\documentclass[twocolumn]{aastex631}
\usepackage{amssymb}
\usepackage{bm}
\usepackage{CJK}
\usepackage{enumitem}
\shortauthors{Huang et al.}
\graphicspath{{./}{figures/}}

\begin{document}
\begin{CJK*}{UTF8}{bsmi}
\title{A Matched Survey for the Enigmatic Low Radio Frequency Transient ILT~J225347+862146}

\author[0000-0003-4267-6108]{Yuping Huang (黃宇平)}
\altaffiliation{LSSTC DSFP Fellow}
\affiliation{Cahill Center for Astronomy and Astrophysics, MC 249-17 California Institute of Technology, Pasadena, CA 91125, USA}
\affiliation{Owens Valley Radio Observatory, California Institute of Technology, 100 Leighton Lane, Big Pine, CA, 93513-0968, USA}
\email{yupinghyper@gmail.com}
\author[0000-0003-2238-2698]{Marin M. Anderson}
\affiliation{Jet Propulsion Laboratory, California Institute of Technology, 4800 Oak Grove Drive, Pasadena, CA 91109, USA}
\affiliation{Cahill Center for Astronomy and Astrophysics, MC 249-17 California Institute of Technology, Pasadena, CA 91125, USA}
\affiliation{Owens Valley Radio Observatory, California Institute of Technology, 100 Leighton Lane, Big Pine, CA, 93513-0968, USA}
\author[0000-0002-7083-4049]{Gregg Hallinan}
\affiliation{Cahill Center for Astronomy and Astrophysics, MC 249-17 California Institute of Technology, Pasadena, CA 91125, USA}
\affiliation{Owens Valley Radio Observatory, California Institute of Technology, 100 Leighton Lane, Big Pine, CA, 93513-0968, USA}
\author{T. Joseph W. Lazio}
\affiliation{Jet Propulsion Laboratory, California Institute of Technology, 4800 Oak Grove Drive, Pasadena, CA 91109, USA}
\author[0000-0003-2783-1608]{Danny C. Price}
\affiliation{International Centre for Radio Astronomy Research, Curtin University, Bentley, WA 6102, Australia}
\affiliation{Department of Astronomy, University of California Berkeley, Berkeley CA 94720, USA}
\author[0000-0003-4531-1745]{Yashvi Sharma}
\affiliation{Cahill Center for Astronomy and Astrophysics, MC 249-17 California Institute of Technology, Pasadena, CA 91125, USA}



\begin{abstract}
    Discovered in 2011 with LOFAR,
    the $15$~Jy low-frequency radio transient ILT~J225347+862146
    heralds a potentially
    prolific population of radio transients at $<100$~MHz.
    However, subsequent transient searches in similar parameter space yielded no
    detections.
    We test the hypothesis that these surveys at comparable sensitivity
    have missed the population due to mismatched survey parameters.
    In particular, the LOFAR survey used only $195$~kHz of bandwidth at $60$~MHz while other surveys
    were at higher frequencies or had wider bandwidth.
    Using $137$~hours of all-sky images from the Owens Valley Radio Observatory Long Wavelength Array
    (OVRO-LWA), we conduct a narrowband transient search at $\sim10$~Jy sensitivity with timescales from $10$~min to
    $1$~day and a bandwidth of $722$~kHz at $60$~MHz.
    To model remaining survey selection effects, we introduce a flexible Bayesian approach for
    inferring transient rates.
    We do not detect any transient
    and find compelling evidence that our non-detection is inconsistent with the detection of ILT~J225347+862146.
    Under the assumption that the transient is astrophysical, we propose two hypotheses that may explain
    our non-detection.
    First, the transient population
    associated with ILT~J225347+862146 may have a low all-sky density and display strong temporal
    clustering.
    Second, ILT~J225347+862146 may be an extreme instance of the fluence distribution,
    of which we revise the surface density estimate at $15$~Jy 
    to $1.1\times 10^{-7}\deg^{-2}$ with a $95\%$ credible interval of
    $(3.5\times10^{-12}, 3.4\times10^{-7})\deg^{-2}$.
    Finally, we find a previously identified object coincident with ILT~J225347+862146
    to be an M dwarf at $420$~pc.
\end{abstract}

\keywords{radiation mechanisms: non-thermal, radio continuum: general, stars: late-type, methods: statistical}

\section{Introduction} \label{sec:intro}
Over the last decade, a new generation of low radio frequency ($\nu \lesssim 300$~MHz; wavelength $\lambda \gtrsim1$~m)
interferometer arrays based on dipoles have emerged.
Dipole arrays simultaneously offer a large effective area ($\sim \lambda^2/4\pi$) as well as
field of view (FOV)
and are thus well suited to synoptic surveys of the time domain sky.
Scientific exploitation of these instruments has been enabled by advances
in processing technology. 
Progress in digital backends \citep[e.g.][]{xgpu, casper} accommodates wider bandwidth
and larger number of dipoles.
New data flagging \citep[e.g.][]{aoflagger, ssins},
calibration \citep[e.g.][]{noordam04, st15}
and imaging \citep[e.g.][]{wsclean, ddfacet, fhd, idg} algorithms
have drastically improved data quality and processing speed.
Dipole-based instruments like the the Long Wavelength Array \citep[LWA;][]{lwa,lwabis}, 
the LOw Frequency ARray \citep[LOFAR;][]{lofar, aartfaac},
the Murchison Widefield Array \citep[MWA;][]{mwa, mwa2},
the Owens Valley Radio Observatory Long Wavelength Array
\citep[OVRO-LWA;][]{ahe+18,eam+18,kgb+15}, and the Square Kilometre Array-Low \citep[SKA-Low;][]{ska} prototype stations
\citep{wsb+17,dbb+20} have carried out increasingly deeper and wider transient surveys.

Low radio frequency transient surveys may probe different populations of transients
than higher frequency (GHz) radio surveys.
At low radio frequencies, synchrotron-powered incoherent extragalactic transient sources often
evolve on years to decades timescales and are often
obscured by self-absorption \citep{mwb15}.
Meanwhile, we expect coherent emission to be more common at low radio frequencies.
The longer wavelength allows a larger volume of electrons to emit in phase and may lead
to stronger emission \citep{melrose17}.
Observationally, some coherent emission mechanisms
prefer low radio frequencies \citep[e.g. electron cyclotron maser emission,][]{treumann06}
or have steep spectra \citep[e.g. pulsars,][]{jsk+18}.
Despite their potential prevalence at low radio frequencies,
the luminosity function for coherent emission sources at low radio frequencies remains poorly characterized.
Initial transient surveys probing timescales of seconds to years at these frequencies
have made significant
progress into the transient rate-flux density phase space,
but the transient populations at these frequencies remain poorly understood compared to higher radio frequencies.

To date, radio transient surveys below $350$~MHz have only yielded
$8$ transient candidates across all timescales, with no populations or definitive multiwavelength
associations identified (see Table~1 of \citealt{ahe+19} for a summary, and \citealt{kws+21} for
an additional candidate). In addition to the rarity of detections, scintillation due to the ionosphere
or near-Earth plasma, typically lasting
a few seconds \citep{kwsr21} to minutes \citep{ahe+19}, also complicates the interpretation of
individual events.
One can identify these events by their spectral features over a wide
bandwidth and their
coincidences with underlying fainter sources.

Of all the low-frequency radio transient detections so far, the \citet{sfb+16} transient,
\object{ILT~J225347+862146}, stands out for a few reasons.
The high flux density, relatively precise localization ($11\arcsec$), and high implied rate
($16^{+61}_{-15}$~sky$^{-1}$day$^{-1}$)
make the transient promising for follow-up observations and searches for the associated population.
The transient was detected during a
$4$ month long LOFAR Low-Band Antennas (LBA) monitoring campaign of the
Northern Celestial Pole (NCP) with irregular time coverage,
totaling $400$~hours of observing time with a snapshot FOV of $175$~deg$^2$. 
The observing bandwidth was $195$~kHz at $60$~MHz.
The transient peaked at $15$--$25$~Jy and evolved on timescales of around $10$ minutes.
The fact that the transient was unresolved on the maximum projected baseline length of $10$~km
and the relatively long duration of the transient argue against a scintillation event
in the near field due to the ionosphere or near-Earth plasma.

The search for the underlying population of ILT~J225347+862146 was one of the goals
of the first non-targeted transient
survey with the OVRO-LWA \citep{ahe+19}. Despite having searched for one order of magnitude larger sky
area than did \citet{sfb+16} at a comparable sensitivity and frequencies,
\citet{ahe+19} reported no detected transients.

One hypothesis that may explain the non-detection by \citet{ahe+19}, which searched in images integrated over
the full $27$--$85$~MHz frequency
coverage of the OVRO-LWA, is that the emission
associated with this transient is confined to a narrow band of frequencies. Coherent transient emission
is known to exhibit narrowband
morphology. Recently, \citet{cpf+21} detected a burst from a M dwarf binary, CR Draconis, that only occupied
a fractional bandwidth of $\Delta\nu/\nu=0.02$ at observing frequency $\nu=170$~MHz.
On the brightest end of coherent emission, Fast Radio Bursts also commonly only appear in a fraction of the
observing bandwidth with typical $\Delta\nu/\nu\sim0.2$ \citep[see e.g.][]{pmb+21},
with an extreme case reaching $\Delta\nu/\nu=0.05$ \citep{ksf+21}.

Motivated by the narrowband hypothesis, the purpose of this work is to search for narrowband transients
with timescales from $10$ minutes to $1$ day in 137 hours of all-sky monitoring data with the OVRO-LWA.
With a comparable bandwidth and sensitivity, we also aim to replicate the
\citet{sfb+16} experiment with two orders of magnitude higher surface area searched.
We also develop a Bayesian model for survey results so that we can fully account for our varying
sensitivity as a function of FOV and robustly assess whether survey results are consistent.

We introduce the OVRO-LWA observation and data collection procedure in
\S~\ref{sec:obsie}.
We describe the visibility flagging and calibration procedures in \S~\ref{sec:cal}, the imaging steps
in \S~\ref{sec:imaging}, and the transient candidate identification pipeline in \S~\ref{sec:tp}.
In \S~\ref{sec:comp}, we introduce a Bayesian approach for modeling transient surveys and
comparing different survey results. \S~\ref{sec:res} details the result of our survey.
In \S~\ref{sec:disc}, we present an M dwarf coincident with the transient ILT~J225347+862146 and discuss
the implications of our work.
We conclude in \S~\ref{sec:conc}.

\begin{deluxetable}{c c }[htb]
\tablecolumns{2}
\tablecaption{Parameters of the Observing Campaign}
\label{tab:obs}
\tablehead{Parameter & Value}
\startdata
Start Time & 2018-03-21~01:28~UTC \\
End Time & 2018-03-26~18:53~UTC \\
Total Observing Time & $137$ hours \\
Maximum Baseline & $1.5$~km \\
Frequency Range & $27.38$--$84.92$~MHz \\
Channel Width & $24$~kHz \\
\enddata
\end{deluxetable}
\section{Observations} \label{sec:obsie}
The OVRO-LWA is a low radio frequency dipole array currently under development at
OVRO in Owens Valley, California. ``Stage II'' of the OVRO-LWA, identical to that in 
\citet{ahe+19}, produced the data for this work.
The final stage of the array will come on-line in 2022, with $352$ antennas spanning $2.4$~km.
The Stage II OVRO-LWA consisted of $256$ dipole antennas spanning a maximum baseline of $1.5$~km.

This transient survey make use of data from a $5$ day observing campaign, the parameters
of which we summarize in Table~\ref{tab:obs}. 
Full cross-correlations across the entire 256-element
array were recorded to enable all-sky imaging.
Stage II of the array only allowed integer second integration time.
As a result, we chose the $13$~s integration time to enable differencing of images
at almost the same sidereal time (see the motivation for sidereal image subtraction in \S~\ref{sec:imaging}),
because $1$ sidereal day is, within $0.1$~s, an integer multiple of $13$~s.
We searched for transients in the $611$~s integrated images (henceforth referred to as the $10$~min search).

Unlike \citet{ahe+19}, which searched for broadband ($\Delta\nu/\nu>1$) counterparts to ILT~J225347+862146,
we explore the possibility that the event was narrowband, with $\Delta \nu / \nu  \ll 1$. 
In our narrowband search, we chose a central frequency of $60$~MHz, identical to that used in \citet{sfb+16}.
\citet{sfb+16} used a bandwidth of 195 kHz, equivalent to  $\Delta \nu/ \nu = 0.003$. In order to ensure
that our sensitivity is well-matched to the peak flux density of ILT~J225347+862146 ($15$--$25$~Jy),
we use a bandwidth that is 3.7 times larger ($722$~kHz) to reach the desired noise level in $10$ min integrated
images.
This decision is well justified because our search is still sensitive to events with $\Delta \nu/ \nu > 0.012$, which is
narrower bandwidth than any known phenomenon discussed in \S~\ref{sec:intro}.
While we only use $722$~kHz of bandwidth for the search, we subsequently incorporate
the full $57.8$~MHz bandwidth for candidate characterization.

\section{Data Reduction and Analyses}
\subsection{Flagging and Calibration}
\label{sec:cal}
\begin{figure}[htb]
    \epsscale{1.5}
    \plotone{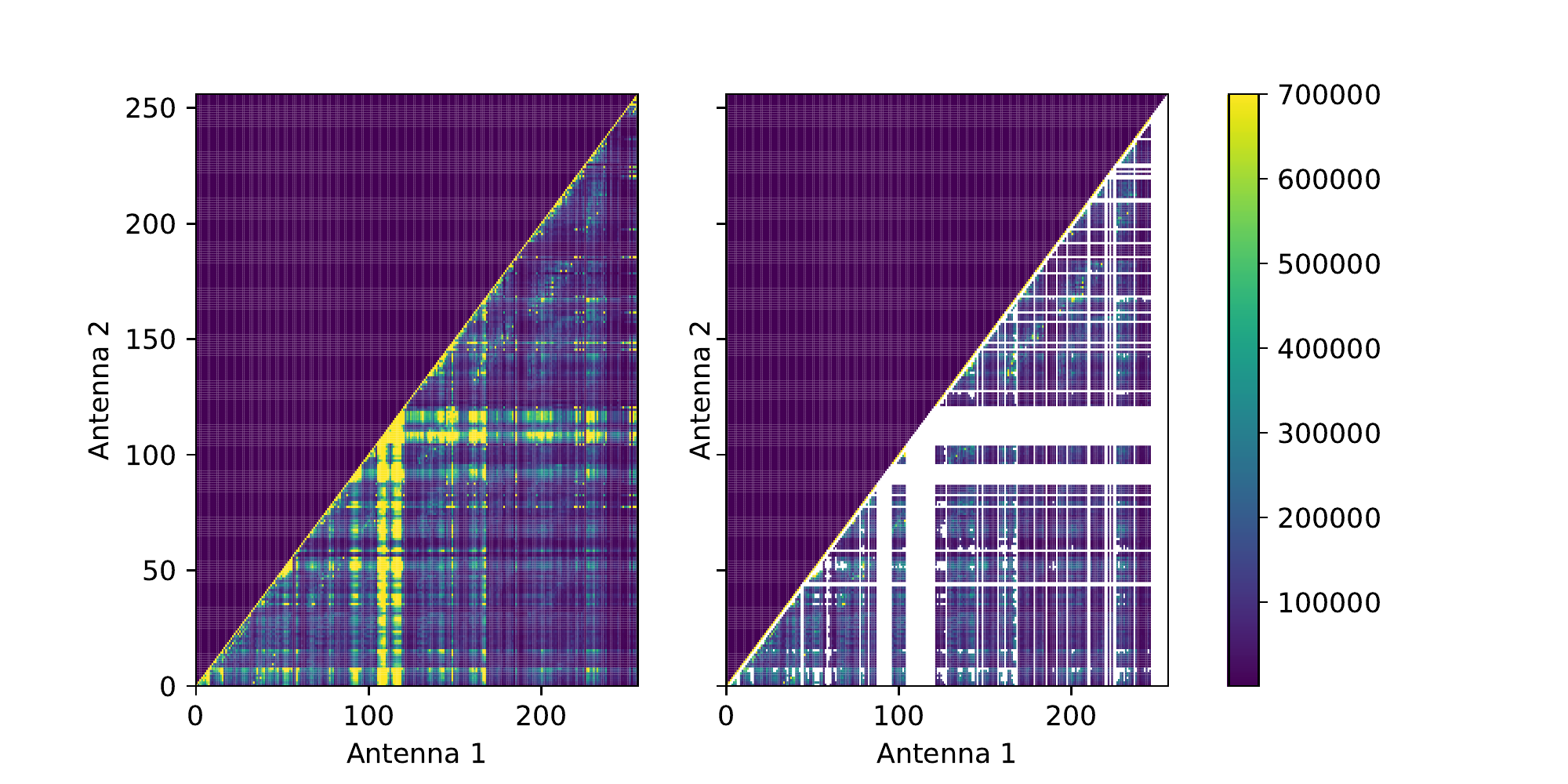}
    \caption{\label{fig:baseline-flag}Amplitude diagnostics for all pairs of baselines before (left) and after (right)
baseline flagging. Due to cross-talk between adjacent signal paths,
a priori flagging of antennas adjacent to each other in the signal path
has been applied before baseline flagging.
The amplitude shown is the frequency-averaged amplitude after time
averaging for $12$ hours without phase tracking. Therefore, outliers indicates bad
antennas or baselines with excess stationary power. The final upgrade of the OVRO-LWA array
will feature redesigned electronics with much better signal paths isolation and thus minimize 
signal coupling between nearby signal paths.}
\end{figure}
Flagging of bad data and calibration for this work largely follow the
procedures outlined in \cite{ahe+19}, which we summarize here.
For each day of observation, we identify and flag bad antennas from their autocorrelation spectra
and derive the direction-independent (bandpass) calibration solutions during Cygnus A transit with the \texttt{bandpass}
task in \texttt{CASA 6} \citep{casa,casa6}.
The bandpass calibration sets the flux scale.
We then apply the daily bandpass solutions and flags to each $13$~s integration for the rest of the day.
For each integration where Cyg A or Cas A are visible, we use
\texttt{TTCal}\footnote{\url{https://github.com/ovro-lwa/TTCal.jl/tree/v0.3.0/}}\citep{TTCal},
which implements the \mbox{StEFCal} algorithm \citep{stefcal}, to solve for
the their associated direction-dependent gains and
and subtract their corrupted visibility from the data, a process known as peeling
\citep{noordam04}.
Peeling solutions are derived once per 13~s integration per 24~kHz frequency channel.
Finally, for each integration, we find bad channels by detecting outliers in averaged visibilities
per channel over baselines longer than 30 meters. The 30-meter cutoff suppresses flux contribution
from the diffuse emission in the sky and allows for more robust outlier detections. The channel flags
are subsequently applied to the $13$~s integration.

Our modifications to the \cite{ahe+19} flagging and calibration approach are as follows:
\begin{enumerate}
    \item \citet{ahe+19} used 13 seconds of data during Cygnus A transit to derive the bandpass calibration.
        In this work, we use $20$ minutes of data around Cygnus A transit.
        The calibration integration time is longer than the typical
        ionospheric and analog gain fluctuation timescales of the array and thus offers more robust
        solutions that are more representative of the instrument bandpass.
    \item To further identify baselines 
        that have excess power due to cross-talk and common-mode noise, we follow \cite{eam+18}'s strategy
        and derive baseline flags by identifying outliers in $12$ hour averaged visibility data without
        phase-tracking after bandpass calibration.
        We pick the 12 hours of the day when the the galaxy is below horizon. Averaging the visibility 
        without phase-tracking attenuates the
        sky signals and highlights stationary excess power on baselines.
        Fig.~\ref{fig:baseline-flag} illustrates this strategy. These flags are generated and applied each day.
    \item For each day, we randomly select two integrations to validate the flags and calibration solutions.
        We identify additional baselines and antennas that show excess visibility amplitude by visual
        inspection and add them
        to the per-day set of flags.
\end{enumerate}
These flagging and calibration steps produce visibility data with flags at $13$~s time resolution.

\begin{figure*}[htb]
    \plotone{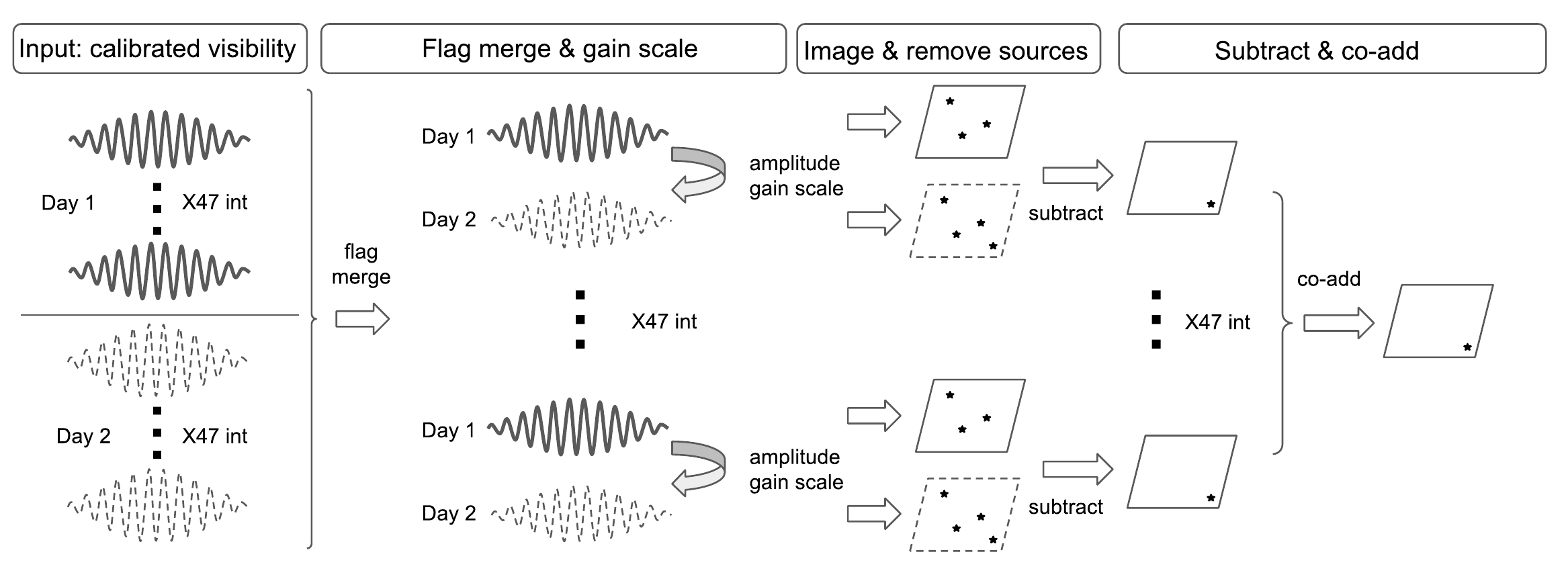}
    \caption{\label{fig:pipeline}A cartoon representation of the imaging and differencing steps that
        produces the differenced images that we search for transients. The inputs are calibrated visibility from
        two time steps being subtracted, separated by one sidereal day.
        Each input visibility integration
        (represented by the fringe pattern) is 13 s long. The group of visibility data from each day
        consists of $47$ integrations. The flag merge, gain scale, imaging,
source removal, subtract, and co-add steps are detailed in \S~\ref{sec:imaging}}
\end{figure*}
\subsection{Imaging and Sidereal Image Differencing}
\label{sec:imaging}
In principle, image differencing allows us to remove diffuse emission and search for
transients below the Jansky-level confusion limit \citep{a04}.
However, when differencing OVRO-LWA images that were a few minutes apart,
\citet{ahe+19} observed the sensitivity degrading compared to the seconds-timescale search.
They concluded that in searches for transients beyond a few integrations,
sources' motions across the antenna beams
introduced significant direction-dependent errors that failed to subtract over the course of
a few minutes.

\begin{figure*}[htb]
    \epsscale{1.0}
    \gridline{
        \fig{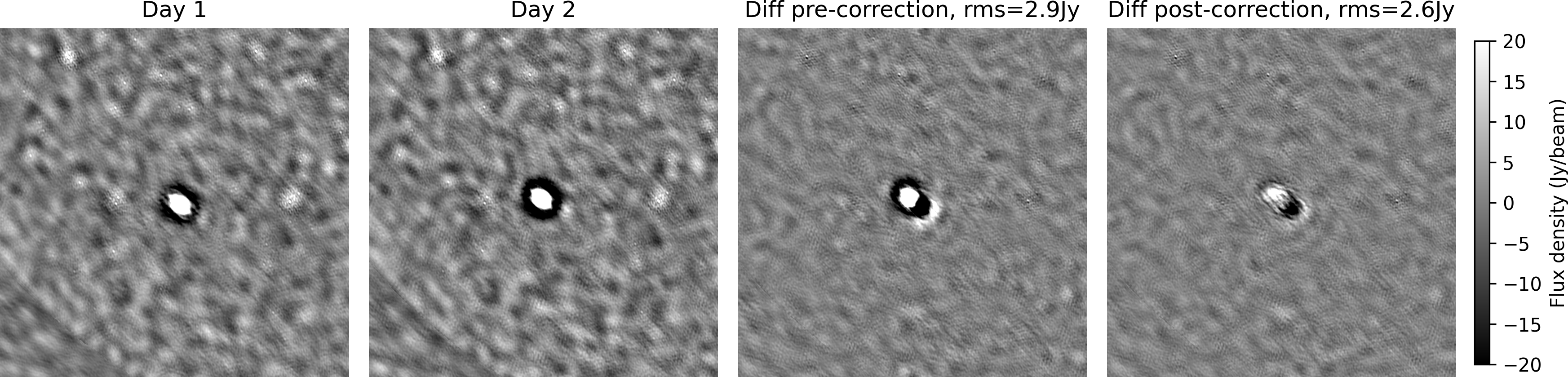}{0.8\textwidth}{(a) $37.5\deg\times37.5\deg$ around the Sun.}
}
    \gridline{
        \fig{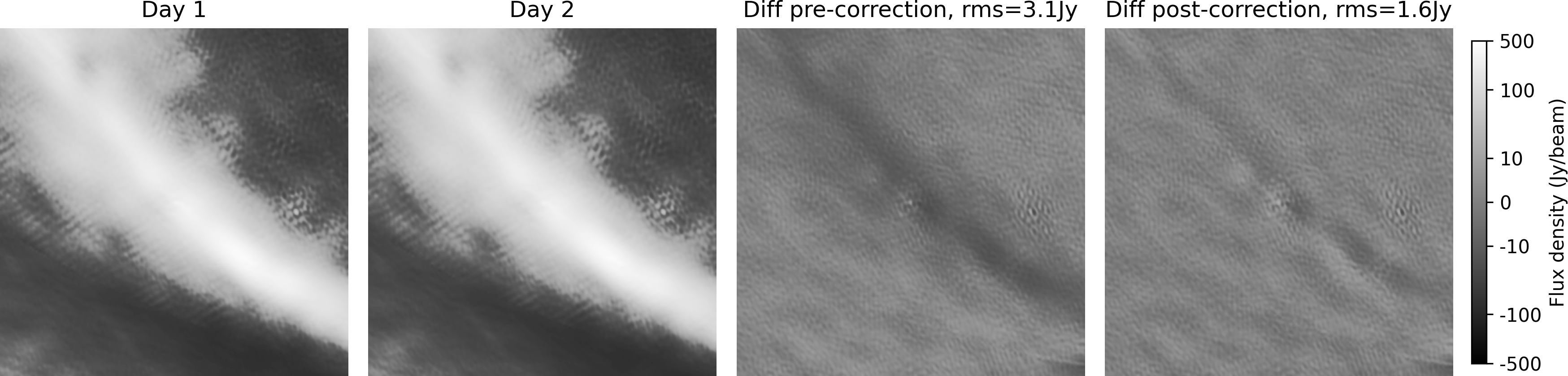}{0.8\textwidth}{(b) $18.75\deg \times 18.75\deg$ around the Galactic Center.}
}
    \caption{\label{fig:sid-art}Images illustrating effects that raise the noise level
        in sidereal image differencing and how we mitigate them. The rms noise is the
        rms noise reported by the source detection code.
    (a) The Sun moves by $\sim1\deg$ per day. Deconvolving the Sun during imaging reduces the
noise due to its sidelobes. (b) The analog gain scaling and inner Tukey weighting suppresses image differencing 
artifacts due to the diffuse sky, especially in the direction of the Galactic plane.}
\end{figure*}

To circumvent the limitations due to the antenna beams, in this work we expand on the sidereal
image differencing technique initiated by \citet{ahe+19}. We difference 
integrations that are, within $0.1$~s, 1 sidereal day apart, so that all persistent sources remain
in the same positions of the antenna beams.
Sidereal image differencing allows clean source subtraction without incorporating the individual
antenna beams into calibration and imaging.
This section details steps for generating
$10$~min integrated and sidereally-differenced images (see also Fig.~\ref{fig:pipeline}). 
For each pair of $10$~min groups of $13$~s visibility data that are 1 sidereal day apart,
we perform the following operations:
\begin{enumerate}
    \item We merge the flags for the two groups and apply the merged flags to all integrations within
        the groups. This ensures that the resultant images for the two groups have the same
        point spread function (PSF).
    \item We apply a per-channel per-antenna per-integration amplitude correction to the integrations from
        the first day so that
        its autocorrelation amplitudes match those from the second day. This corrects for gain amplitude
        variations on short timescales (most notably temperature-dependent analog electronics gain variation
        that correlates with the $15$ min air-conditioning cycle in the electronics shelter). 
    \item We change the phase center of all visibility data to the same sky location, the phase center
        in the middle of the time integration. We then image each $13$ s
        integration with \texttt{wsclean} \citep{wsclean}, using Briggs~$0$ weighting and a inner
        Tukey tapering parameter (\texttt{-taper-inner-tukey}) of 20~$\lambda$. The weighting and tapering scheme
        suppresses diffuse emission, especially toward the galactic plane, without introducing ripple-like
        artifacts corresponding to a sharp spatial scale cutoff. The typical
        full width at half maximum (FWHM) of the synthesized beam is $23'\times13'$.
    \item During imaging, we allow deconvolution of the Sun and the Crab pulsar by masking everything else
        in the sky with the \texttt{-fits-mask} argument of \texttt{wsclean}. We set the
        \texttt{CLEAN} threshold to $50$~Jy. This removes sidelobes in the images due to the Sun and the
        Crab pulsar: the Sun moves in celestial coordinates from day to day,
        and the Crab pulsar exhibits strong variability.
    \item Each image from the first day is subtracted from its sidereal counterpart from the second day to
        form the differenced image.
        We then co-add the group of differenced images to form the $10$~min
        differenced image.  We chose the co-adding approach because it is more
        efficient to parallelize than gridding all 10~minutes of visibility.
        For a subset of our data, we confirm that
        the co-added differenced images suffer from no sensitivity loss or
        artifacts by comparing them
        to differenced images produced directly by imaging the full $10$~min visibility dataset.
\end{enumerate}

\begin{figure*}[htb]
    \gridline{
    \fig{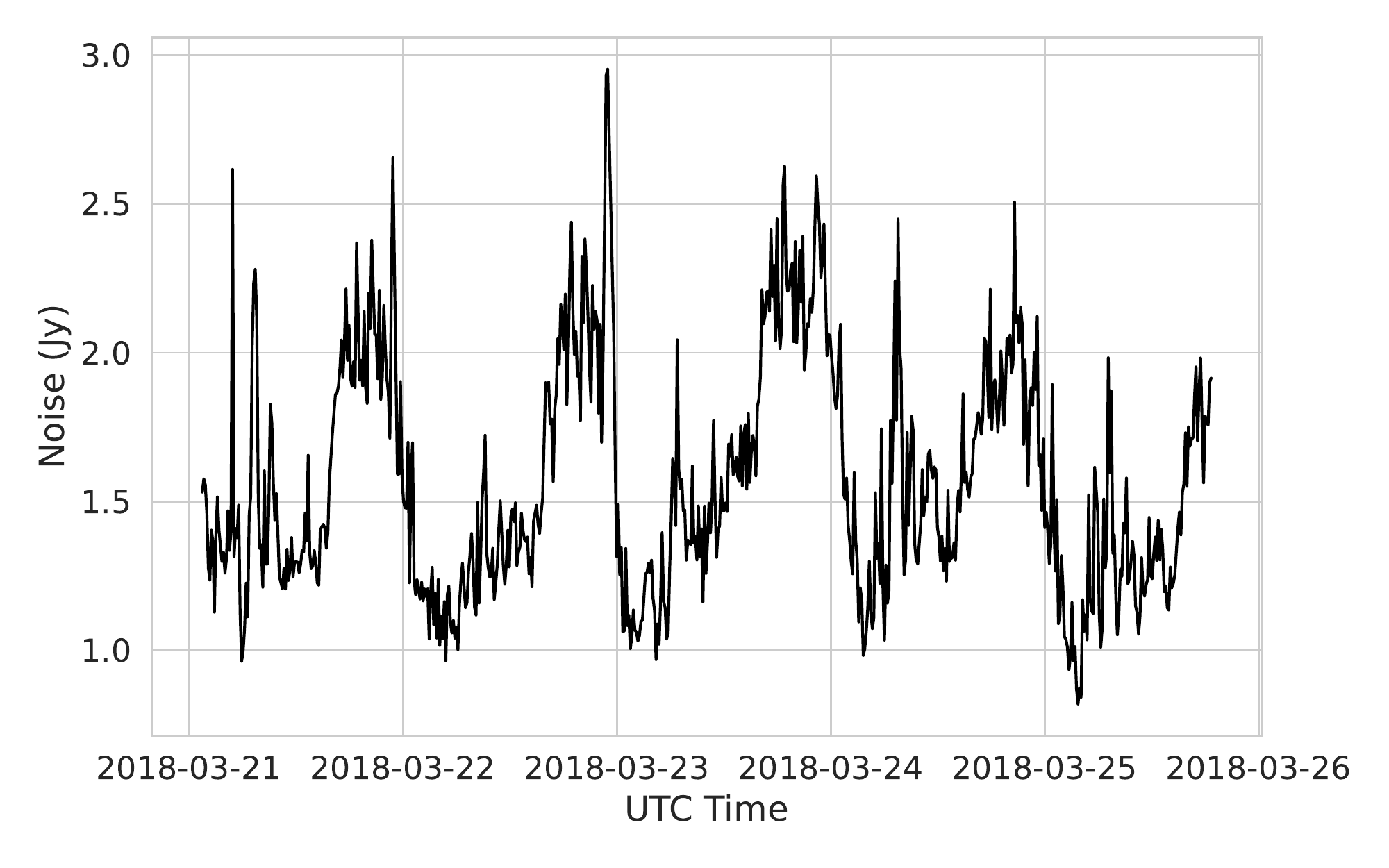}{0.5\textwidth}{(a)}
    \fig{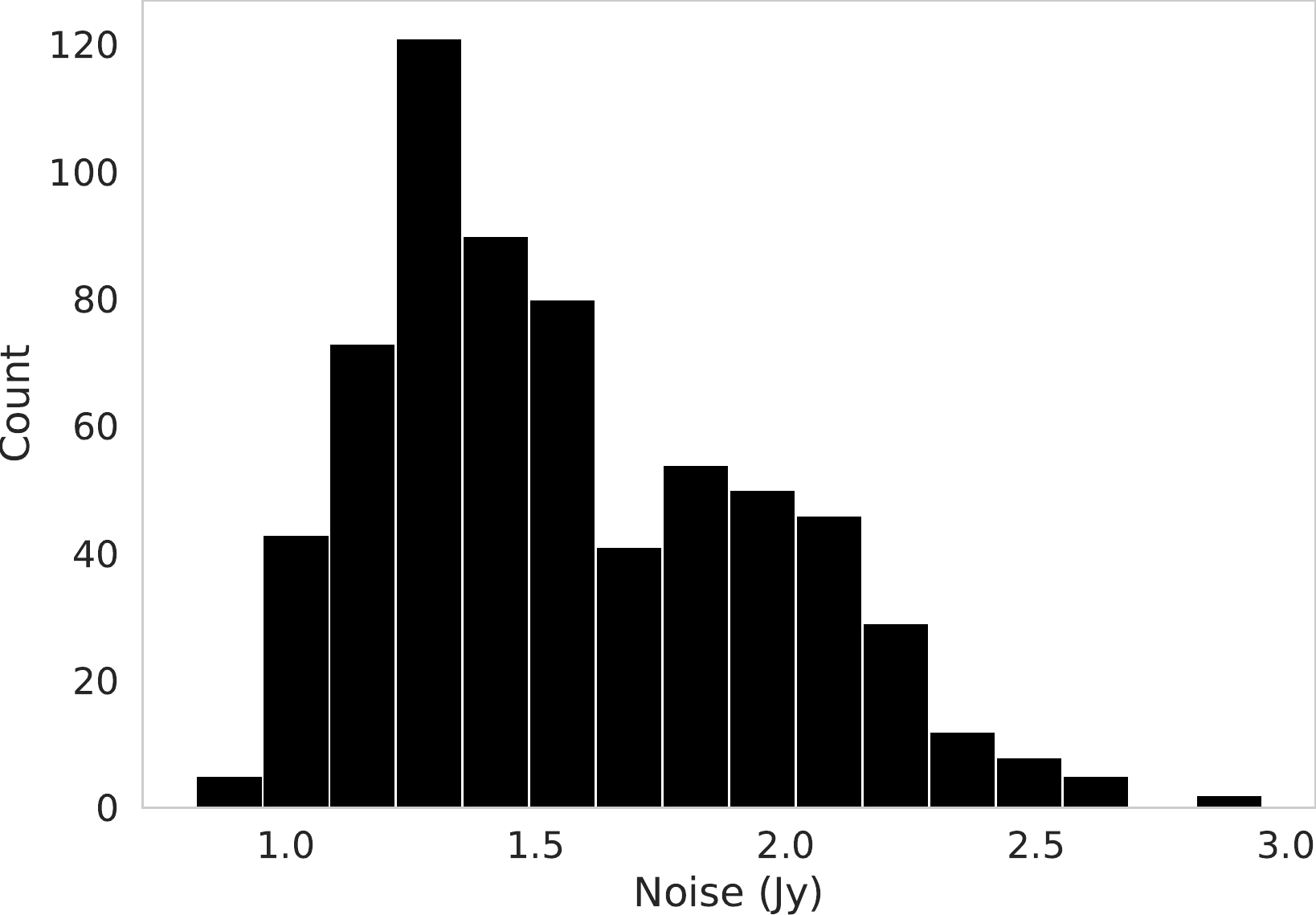}{0.43\textwidth}{(b)}
}
\caption{\label{fig:noises} \textbf{(a)}~Time series of noise at zenith in $10$~min subtracted images over the
    entire observation. Higher noise level
    corresponds to daytime. Noise level spikes typically occur at sunrise, at sunset,
    when a horizontal RFI source flares up, and when the Crab pulsar scintillates.
    \textbf{(b)}~Histogram of image-plane noise measured in
    all integrations. The two modes
    of the distribution correspond to daytime (when both the Sun and the galactic plane are up)
and nighttime observations.}
\end{figure*}

Fig.~\ref{fig:sid-art} shows the main classes of problematic image differencing artifacts
that our procedure removes. Our procedure aims at reducing the root-mean-square (rms) estimate of the
noise due to far sidelobes
of these artifacts in the rest of the image. The sidereally
differenced images that our procedure produce are
the data product on which we perform source detection to search for transients.
Fig.~\ref{fig:noises} shows the noise characteristics of the sidereally differenced images.

We use \texttt{Celery}\footnote{\url{https://docs.celeryproject.org/en/stable/}}, a distributed task queue framework,
with \texttt{RabbitMQ}\footnote{\url{https://www.rabbitmq.com/}} as the message broker to distribute
    the compute workload for this project across a 10-node compute cluster near
    the telescope. Each node has $16$ cores and $64$~GB of RAM. The snapshot of the pipeline source code 
used for this work can be found at \url{https://github.com/ovro-lwa/distributed-pipeline/tree/v0.1.0}.
\subsection{Source-finding and Candidate Sifting} \label{sec:tp}

We use the source detection
code\footnote{\url{https://github.com/ovro-lwa/distributed-pipeline/blob/v0.1.0/orca/extra/source_find.py}} developed by \citet{ahe+19}
to detect sources in the sidereally subtracted images. The algorithm divides each image into $16$ tiles and
estimates the local image noise in each tile. It then
groups bright pixels with a Hierarchical Agglomerative Clustering (HAC) algorithm to identify individual
sources.
\citet{ahe+19} tuned the parameters of the HAC algorithm for detecting sources in dirty subtracted images
of the OVRO-LWA.
The source detection algorithm only reports sources with peak flux density $6.5$ times
the local standard deviation $\sigma$.
Based on the number of independent
synthesized beam searched \citep{fko+18}, we estimate the probability of
detecting a $6.5\sigma$ outlier due to Gaussian noise fluctuation
over the entire survey to be $<5\times 10^{-3}$.

For each detected source, we visually inspect its cutout images and its all-sky image in an interactive \texttt{Jupyter}
        \citep{jupyter}
notebook widget\footnote{\url{https://github.com/ovro-lwa/distributed-pipeline/blob/v0.1.0/orca/extra/sifting.py}}
that records the labels for all detected
sources.
We developed the tool with the
\texttt{ipywidgets}\footnote{\url{https://github.com/jupyter-widgets/ipywidgets}}
and
\texttt{matplotlib} \citep{matplotlib} packages.
We can rule out a large number of artifacts based on
their appearances and their positions in the sky: RFI sources and meteor reflections
are often resolved and/or close to the horizon.
We label point sources detected in the subtracted images that only appear
in either the ``before'' or the ``after'' images as candidate transients.

For these candidates, we generate spectra time series (dynamic spectrum) over the entire
$58$~MHz of bandwidth and re-image them with different weighting schemes to
ascertain the properties of these candidates. For candidates that appear
near Vir~A, Tau~A, or Her~A, we deconvolve the bright source to test
whether a given candidate is part of the bright source's sidelobe.

\subsection{Quantifying Survey Sensitivity}
\label{sec:quant}
We quantify the noise in subtracted images with the standard deviations at zenith
reported by the source detection code.

The power beam of an OVRO-LWA dipole approximately
follows a $\cos^{1.6}(\theta)$ pattern, where $\theta$ is the
angle from zenith \citep{ahe+19}. Therefore, for a given snapshot with noise at zenith $\sigma_z$, the
primary-beam-corrected image noise
at an angle $\theta$ from zenith is given by $\sigma_z/\cos^{1.6}(\theta)$.
Furthermore,
the number of artifacts increases as the zenith angle increases, due to both horizon RFI sources and
increased total electron content (TEC) through the ionosphere at lower elevations.
Therefore, we define the zenith angle cutoff for our survey as when the marginal volume probed
with increasing zenith angle is small.
The volume probed for a non-evolving population of transients 
uniformly distributed in space has the following dependencies on FOV and sensitivity:
\begin{equation}
    V \propto \int_{0}^{\theta_0} S_0^{-3/2}d\Omega,
\end{equation}
where $S_0$ is the sensitivity as a function of solid angle $\Omega$, and $\theta_0$ the
zenith angle limit of a survey. This is equivalent to the Figure of Merit defined in
\cite{macquart14} for such a population of transients. Substitute in
the dependency of sensitivity on zenith angle and we get
\begin{eqnarray}
    V &\propto& \int_{0}^{\theta_0} (\cos^{-1.6}{\theta})^{-3/2} \sin{\theta} d\theta \nonumber \\
    &\propto& -\cos^{3.4}\theta_0 \nonumber.
\end{eqnarray}
We choose a zenith angle cut
$\theta_0=60\deg$, which encompasses $90\%$ of the available survey volume. The beam-averaged
noise $\bar{\sigma}$ is therefore given by
\begin{equation}
    \label{eq:sens}
    \bar{\sigma }
    = \frac{\int_0^{2\pi}\int_0^{\theta_0} \frac{\sigma_z}{\cos^{1.6}\theta}\sin\theta d\theta d\phi}{\int_{0}^{2\pi} \int_{0}^{\theta_0}\sin\theta d\theta d\phi}.
\end{equation}
For a zenith angle cut of $\theta_0=60\deg$, this evaluates to $1.72\sigma_z$.

\begin{figure}
    \epsscale{1.1}
    \plotone{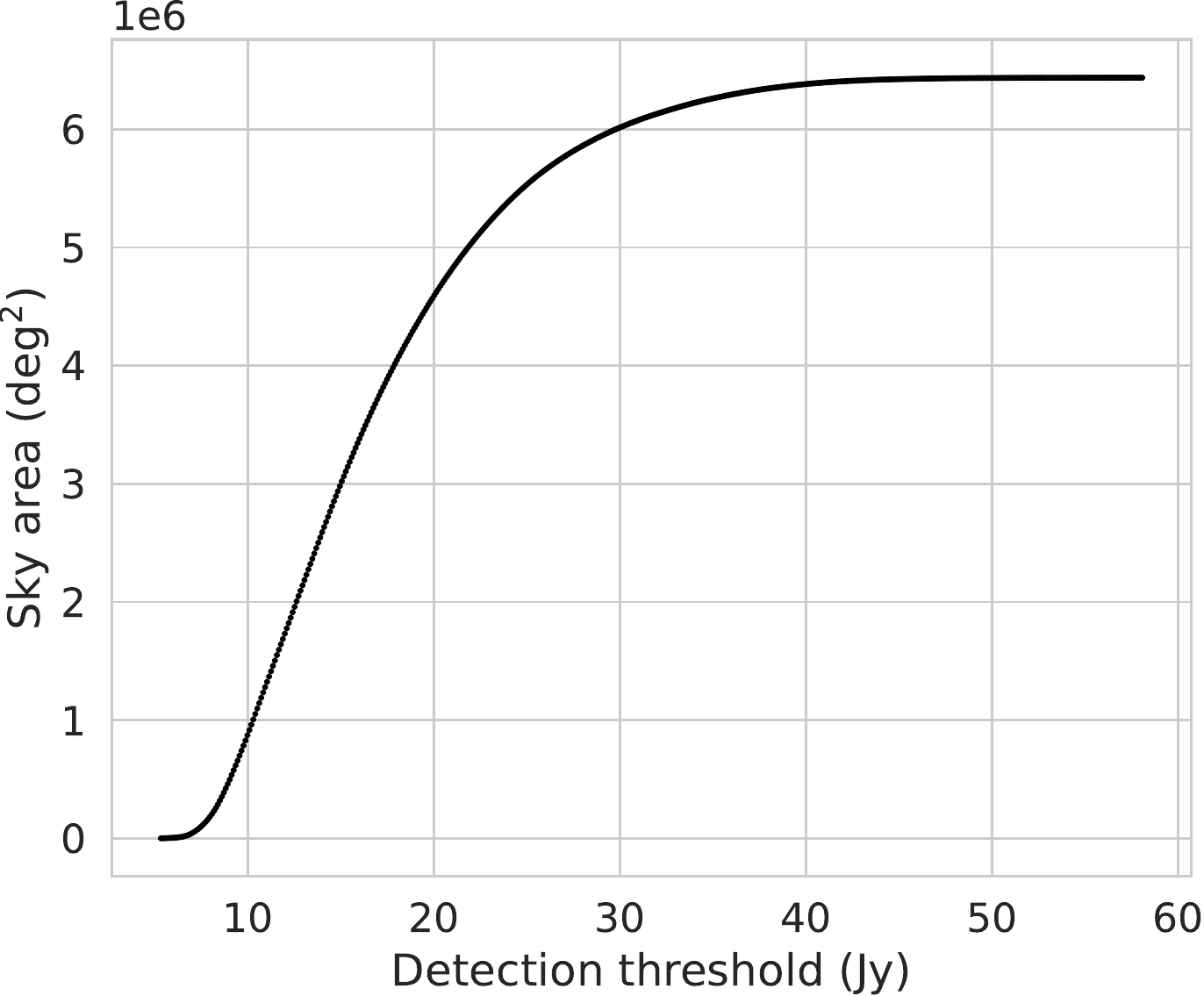}
    \caption{\label{fig:rho-s}Cumulative sky area surveyed at $10$~min timescale as a function of detection threshold.}
\end{figure}

Since our sensitivity varies significantly over the FOV, we also quantify our
sensitivity in terms of total sky area versus sensitivity, aggregated over all images
in our survey. Our approach is similar to that of \citet{bmk+14}, albeit with much finer
flux density bins. Fig.~\ref{fig:rho-s} shows the cumulative sky area as a function
of sensitivity for $10$~min timescale transients. The binned sky area and sensitivity
$\{\Omega_{tot, i}, S_i\}$ forms the basis of our Bayesian modeling of transient detections
detailed in \S~\ref{sec:bayes}.

The aforementioned approach assumes that the sky is static with respect to the primary beam.
    However, Earth rotation rotates the sky across the primary beam.
    We do not account for for this effect in our analysis due to the short integration time and the
    smoothness of the primary beam.
    The rotation modifies the sensitivity estimate for each point in the sky by a negligible $<1\%$ for a 10 min integration.

\section{Estimating the Transient Surface Density}
\label{sec:comp}
While our survey aims to match \citet{sfb+16} as much as possible, there remains
a number of differences. Most notably, our sensitivity varies by factor of $\sim 8$
across the survey, due to the gain pattern of a dipole antenna and
different level of sky noise at different time of the day. Therefore, in this
section, we devise a Bayesian scheme for inferring transient rates so that we can 
incorporate varying sensitivity as a function of sky area surveyed. The Bayesian
approach also facilitates testing whether two survey results are consistent,
an important question when the implied rate of two surveys are significantly different.

\subsection{The Frequentist Confidence Interval}
\begin{figure*}
    \epsscale{1.0}
    \plotone{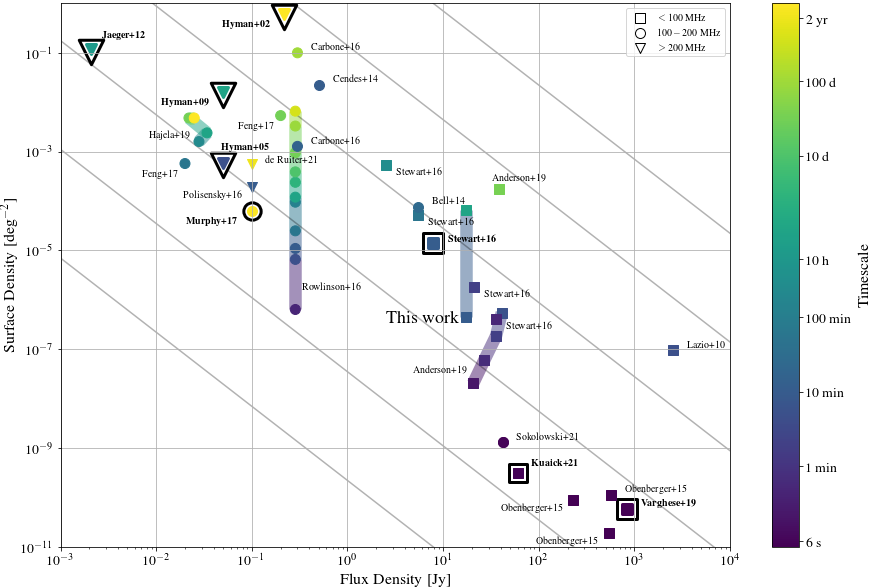}
    \caption{\label{fig:phasespace}
    The radio transient phase space diagram shows the transient surface density as a function
    of limiting flux density for non-targeted transient surveys at $<300$~MHz to date.
    Each point denotes the typical sensitivity and the $95\%$ frequentist upper limit of transient
    surface density of the survey.
    Surveys with detections are marked in bold.
    The color denotes the timescale of the search, ranging from timescales of $1$~s \citep{kws+21}
    to $5.5$~-yr \citep{rlr+21}. Surveys conducted at different frequencies are marked with
    different shapes. 
    Surveys with similar surface density and flux density limits may probe different
    populations of transients if they operate in different frequencies or timescales.
    Each of the solid gray lines traces a hypothetical standard candle population in a
    Euclidean universe, i.e. a cumulative flux density distribution
    (Eq.~\ref{eq:rho}) power law index of $\gamma=3/2$.\\
    References: \citet{hlk+02, hlk+05, hwl+09, lcl+10, jhk+12, bmk+14,
    cws+14, oth+15, chw+16, plh+16, rbm+16, sfb+16,fvh+17,mkc+17, ahe+19, hmi+19, vbd+19,
kws+21, rlr+21, swb+21}.}
\end{figure*}

\label{sec:freq}
Once we count the number of transients $n$ detected in a survey, we can estimate
the rate of low-frequency transients. For a given timescale, the rate of transients
above a certain flux density threshold $S_0$ is typically parameterized by the surface
density $\rho$, which gives the number of transients per sky area.
For a given population of transient
that occur with a surface density $\rho$ above a certain flux threshold $S_0$,
the number of detections in a given survey
with total independent sky area surveyed $\Omega_{tot}$ follows a Poisson distribution with
rate parameter 
\begin{equation}
    \label{eq:lamb}
    \lambda=\rho\Omega_{tot}.
\end{equation}
The probability mass function (PMF) of the
Poisson distribution is given by
\begin{equation}
    \label{eq:pois}
    P_{pois}(n|\lambda) = \frac{\lambda^n e^{-\lambda}}{n!},
\end{equation}
where $P(n)$ is the probability of obtaining $n$ detections.
\citet{g86} computed a table of confidence interval values for $\lambda$ for a range of probability
and number of detections in a given survey, from which one can derive the confidence
interval on the surface density $\rho$.
The 95\% upper limit on the surface density $\rho$, along with the survey sensitivity $S_0$,
is the typical metric quoted
in low-frequency radio transient surveys and are plotted in the phase space diagram
(Fig.~\ref{fig:phasespace}). 

Our survey is sensitive to transients with decoherence timescale \citep{macquart14} $T$
from $10$ minutes to $1$ day.
Since each
of our snapshot has the same FOV $\Omega_{FOV}$, the total independent sky area surveyed is given
by
\begin{equation}
    \label{eq:omegatot}
    \Omega_{tot} \simeq \Omega_{FOV}\left\lfloor\frac{N}{T/10\text{min}}\right\rfloor,
\end{equation}
where $N$ is the
number of $10$~min sidereally differenced images and $\lfloor \cdot \rfloor$ the
floor function. Following conventions in the low-frequency transient search literature,
we quote the $95\%$ confidence upper limit on $\rho$ at the average
sensitivity of the survey.

\subsection{Bayesian Inference for Transient Surveys}
\label{sec:bayes}
For wide-field instruments at low frequencies, the survey sensitivity 
can vary by more than an order of magnitude with time and FOV.
Different sensitivity probes a different depth for a given population of
transients.
By reducing the information contained in a survey to its typical sensitivity,
the above approach does not use all information contained within a survey. 
To address the variation of sensitivity
across a survey, \citet{chw+16} models the surface density $\rho$
above a flux threshold $S_0$ as a power law of sensitivity:
\begin{equation}
    \label{eq:rho}
    \rho(S>S_0) = \rho_* \left(\frac{S_0}{S_*}\right)^{-\gamma},
\end{equation}
where $\gamma$ is the power law index, and $\rho_*$ the reference surface density at flux density $S_*$.
The Poisson rate parameter is then given by
\begin{equation}
    \label{eq:lamb2}
    \lambda = \rho_* \left(\frac{S_0}{S_*}\right)^{-\gamma}\Omega_{tot},
\end{equation}
For a given $\gamma$, 
the reference surface density $\rho_*$ can be inferred from number of detections in parts of
the survey with different sensitivity.

Here we develop a Bayesian approach that extends the \citet{chw+16} model.
Apart from enabling future extensions to the model, the main utilities of the Bayesian approach are as follows:
\begin{enumerate} 
    \item it allows us to marginalize over
        the source count power law index $\gamma$ for an unknown population
        when inferring the surface density $\rho_*$;
    \item it outputs posterior distribution
        over $\rho_*$, which can be integrated to inform future survey decision making;
    \item it allows for robust hypothesis testing of whether survey results are consistent
        with each other.
\end{enumerate}

Our baseline model, $\mathcal{M}_1$, jointly infers $\gamma$ and $\rho_*$ for a single
population of transients,
thereby naturally accommodating our survey's change of surface area with sensitivity.
The alternative model, $\mathcal{M}_2$, proposes that our survey probes
a population with surface density $r\rho_*$, with $r$ as a free parameter.
In other words, $\mathcal{M}_2$ proposes that our survey and
\citet{sfb+16} select for different population of transients.
Model comparison between $\mathcal{M}_1$ and $\mathcal{M}_2$ informs us whether
two transient surveys yield inconsistent results. We now elaborate on the details
of the models. The notebooks that implement the models are hosted at
\url{https://github.com/yupinghuang/BIRTS}.

\subsubsection{The Setting}
To infer the model parameters $\bm{\theta}$ for a given model $\mathcal{M}$ and measured
data $D$, we use Bayes' theorem to obtain the posterior distribution, the probability distribution 
of $\bm{\theta}$ given the data,
\begin{equation}
    \label{eq:fave}
    p(\bm{\theta}|D,\mathcal{M}) =
    \frac{p(D|\bm{\theta},\mathcal{M})p(\bm{\theta}|\mathcal{M})}{p(D|\mathcal{M})}.
\end{equation}
Several other probability distributions of interest appear in Bayes' theorem.
$p(D|\bm{\theta},\mathcal{M})$ is the likelihood function, the probability of
obtaining the measured data $D$ given
a fixed model parameter vector $\bm{\theta}$ under model $\mathcal{M}$.
$p(\bm{\theta}|\mathcal{M})$ is the prior distribution, specifying our a priori belief
about the parameters. $p(D|\mathcal{M})$ is the evidence, the likelihood of observing data $D$ under
model $\mathcal{M}$. Normalization of probability to $1$ requires that
\begin{equation}
    \label{eq:fave2}
    p(D|\mathcal{M}) = \int p(D|\bm{\theta}, \mathcal{M}) p (\bm{\theta}|\mathcal{M})
    d\bm{\theta},
\end{equation}
which gives the evidence $p(D|\mathcal{M})$ the interpretation of the likelihood of observing data $p(D)$
averaged over the model parameter space. 

\subsubsection{Representing Data}
We encode the results of surveys in the data variable $\{D_i\}=\{S_{0,i}, \Omega_{tot, i}, n_i\}$, where
${S_{0,i}}$ are the sensitivity bins, $\Omega_{tot,i}$ the differential total
area surveyed in the $i$-th bin, and $n_i$ the number of detections in the $i$-th bin. 
The \citet{sfb+16} detection with LOFAR can then be written as a one-bin data point:
\begin{equation}
    \label{eq:dl}
D_{L} = \{15~\text{Jy}, 3.3\times10^5~\deg^2, 1\}.
\end{equation}
For the OVRO-LWA, $\{S_{0,i}, \Omega_{tot,i}\}$ is the differential sensitivity-sky area curve described
in \S~\ref{sec:quant}.
\subsubsection{A Single Population Model}
For a single survey, or for multiple surveys where we assume that the selection criteria
do not affect the  observed rate of the transients, a Poisson model with a single
reference surface density $\rho_*$ and source count power law index $\gamma$ is appropriate.
We denote this model $\mathcal{M}_1$ and the parameters $\bm{\theta_1} = (\rho_*,\gamma)$.

For all the survey data encoded in $\{D_i\}$, the model states
that for each sensitivity bin $S_{0,i}$ with sky area $\Omega_{tot,i}$, the detection count $n_i$
follows a Poisson distribution
\begin{equation}
    \label{eq:m1}
    \mathcal{M}_1: n_i \sim P_{pois}\left(n_i|\lambda=\rho_*\left(\frac{S_{0,i}}{S_*}\right)^{-\gamma}\Omega_{tot,i}\right),
\end{equation}
where we use the $\sim$ operator to denote that each $n_i$ independently follows
the distribution specified by the Poisson PMF
$P_{pois}$ defined in Eq.~\ref{eq:pois}. We choose the reference flux density $S_*=15$~Jy.

With the model specified, we adopt uninformative prior distributions 
$p(\gamma)\propto \gamma^{-3/2}$ and $p(\rho_*)\propto 1/\rho_*$
derived in Appendix~\ref{sec:prior}.
Integrating the joint posterior distribution $p(\rho_*,\gamma|D, \mathcal{M}_1)$ gives the marginalized
posterior distribution for $\rho_*$. To understand the sensitivity of
the posterior distribution on the choice of prior distributions, we also derive 
the posterior with uniform priors on $\gamma$ and $\rho_*$. In all cases,
we bound the prior distribution on on $\gamma$ to $(0,5)$ and
on $\rho_*$ to be $(10^{-14}, 10^{-3})\deg^{-2}$.

Even though the Poisson distribution can be
integrated analytically over $\lambda$, with our modifications the likelihood function
cannot be integrated analytically.
For this two-parameter model, the integral can be done by a Riemann sum over a grid.
However, we adopt a Markov Chain Monte Carlo (MCMC) approach to integrate
the posterior distribution.
The MCMC approach allows
extensions of the model. For example, one may wish to incorporate an upper flux density cutoff
$F_{max}$, for the flux density distribution. We extend this model to test the consistency
of different survey results in the next section.
The MCMC approach
will also allow future work to turn more realistic models for transient detections \citep[see e.g.][and references within]{chw+17, ttw+13}
into inference
problems, which will enable more accurate characterizations of the transient sky.

We use the No-U-Turn Sampler \citep[NUTS;][]{nuts},
an efficient variant of the Hamiltonian
Monte Carlo \citep[HMC;][]{hmc} implemented in the Bayesian inference package
\texttt{pymc3} \citep{pymc3}
to sample from the posterior distribution.
We allow 5000 tuning steps for the NUTS sampler to adapt its parameters and run 4 chains at different 
    starting points. We check the effective sample size and the $\hat{R}$ statstics \citep{vgs+21}
    provided by \texttt{pymc3} for
convergence of the samples to the posterior distribution.

\subsubsection{A Two-population Model}
To answer whether our survey results are consistent with \citet{sfb+16}, we develop
a second model $\mathcal{M}_2$ as the competing hypothesis. $\mathcal{M}_2$ states
that the transient counts from our survey with the OVRO-LWA, $\{n_i\}_{O}$, are drawn from
a different Poisson distribution from which the LOFAR counts $\{n_i\}_L$ are drawn from.
We introduce the surface density ratio, $r$,
which modifies the effective transient surface density $\rho_*$ for our survey.
In other words, $\mathcal{M}_2$ posits that our survey probes a population with a different
surface density $r\rho_*$, than did \citet{sfb+16}.
The model can be written as
\begin{eqnarray}
    \label{eq:m2}
    &&\mathcal{M}_2: \nonumber \\
    &&\{n_i\}_{L} \sim P_{pois}\left(n_i|\lambda=\rho_*\left(\frac{S_{0,i}}{S_*}\right)^{-\gamma}\Omega_{tot,i}\right),\nonumber\\
    &&\{n_i\}_{O} \sim P_{pois}\left(n_i|\lambda=r\rho_*\left(\frac{S_{0,i}}{S_*}\right)^{-\gamma}\Omega_{tot,i}\right).
\end{eqnarray}
Our physical interpretation of $\mathcal{M}_2$ is that the two surveys probe populations with
different averaged transient surface density.

The parametrization with the surface density ratio
$r$ captures a wide range of selection effects, which may result in
different specifications of the prior distribution on $r$.
Since our survey covers the galactic plane,
our all-sky rate can be enhanced if the population is concentrated along the galactic plane.
We speculate that a natural prior on $r$ is then a uniform prior.
On the other hand, the time sampling of
\citet{sfb+16} extends over $4$ months, while we have a continuous $5$ day survey.
If the decoherence timescale
of the transient event is much longer than the $10$~min emission timescale
(e.g. long-term activity cycles), it reduces the number of epochs and thus the
effective total area $\Omega_{tot}$ for our survey. In this case, a uniform prior on $1/r$
might be more appropriate.
Lacking compelling evidence, we do not assume a particular source of rate
modification and prefer the uninformative prior $p(r)\propto 1/r$ derived in Appendix~\ref{sec:prior},
which is invariant under the reparameterization $r\rightarrow 1/r$.
Finally, we can put an additional constraint of $r>1$ or $r<1$ on the prior 
depending on whether we are interested in testing the effective surface density
in our survey is enriched or diluted.

This parameterization, however, does not capture narrow bandwidth of the signal,
because a narrow bandwidth modifies
the effective flux of the transient, which appears inside the exponentiation by $\gamma$
in Eq.~\ref{eq:lamb2}.
Since we explicitly search for narrowband transients (\S~\ref{sec:obsie}),
we do not consider such a model.

\subsubsection{Testing Survey Consistencies via Model Comparison}
With the two models we developed, the question of whether two survey
results are inconsistent translates to deciding which model is preferred
given the data.
Given the dearth of information contained in surveys with few or no
detections, a particular class of methods may inadvertently bias the result.
Therefore, we test three different methods for Bayesian model
comparisons as outlined below and compare their results.
\paragraph{WAIC} 
The first class is based on estimating the
predictive accuracy of models. One popular example is the Widely Applicable Information Criterion
\citep[WAIC;][]{waic, vgb+15}, which can be easily computed from posterior samples.
Given $S$ samples of the parameters $\bm{\theta_s}$ from the computed posterior and all the data
$y_i$, the WAIC is given by
\begin{eqnarray}
    \text{WAIC} = &\sum_{i=1}^n&\log\left(\frac{1}{S} \sum_{s=1}^S p(y_i|\bm{\theta_s}) \right) -\\
    &\sum_{i=1}^n&\text{Var}_{s=1}^S(\log p(y_i|\bm{\theta_s})),
\end{eqnarray}
where $\text{Var}_{s=1}^S$ denotes variance taken over the posterior samples.
The first term is an estimate of the expected predictive accuracy of the model, while the
second term, the effective degree of freedom, penalizes more complex models that are overfitted.
The difference in the WAIC between two models, $\Delta\text{WAIC}$, then gives a measure of
how well the two models may predict out-of-sample data.
\paragraph{Bayes factor}
The second class of model comparison method bases on the Bayesian evidence Eq.~\ref{eq:fave2}, i.e. how
efficient does a model explain observed data. Between two models, one computes the Bayes factor
\begin{equation}
    B_{12} = \frac{p(\mathcal{M}_1|D)p(\mathcal{M}_1)}{p(\mathcal{M}_2|D)p(\mathcal{M}_2)},
\end{equation}
where $p(\mathcal{M}_1)$, and $p(\mathcal{M}_2)$ are the prior distributions on each model, usually taken
to be equal when no model is preferred a priori.
Models with a larger parameter space is penalized by the
resultant lower prior density.
Scales exist for interpreting the significance
of Bayes factor \citep{bf1}. 

\paragraph{Mixture model}
The third method advocates for the use of a mixture model of the two contesting models in question
and basing model comparison off the posterior of the mixture parameter \citep{kmr+14}.
The mixture approach avoids the computational cost and some theoretical difficulties
of the Bayes factor.
To construct the mixture model, we refer to the distribution function that
generates the data under $\mathcal{M}_1$ as $f_1$,
and the distribution function that corresponds to $\mathcal{M}_2$ as $f_2$, such that
Eq.~\ref{eq:m1} is equivalently $\mathcal{M}_1: n_i\sim f_1$, and Eq.~\ref{eq:m2}
is $\mathcal{M}_2: n_i \sim f_2$.
With a parameter $\alpha$ that denotes the mixture weight for model $\mathcal{M}_2$,
$0\leq \alpha \leq 1$.
We construct the mixture model $\mathcal{M}_m$ from $\mathcal{M}_1$ and $\mathcal{M}_2$
for the purpose of model comparison.
$\mathcal{M}_m$ is given by
\begin{eqnarray}
    \mathcal{M}_m: n_i \sim (1-\alpha) &f_1&(n_i|\bm{\theta_1}, \Omega_{tot,i}, S_{0,i}) + \nonumber\\
    \alpha &f_2&(n_i|\bm{\theta_2}, \Omega_{tot,i}, S_{0,i}).
\end{eqnarray}
The mixture weight, $\alpha$, can be interpreted as the propensity
of the data to support $\mathcal{M}_2$ versus $\mathcal{M}_1$.
If $\alpha\rightarrow 1$, then $\mathcal{M}_2$ generates the data.
If $\alpha\rightarrow 0$, $\mathcal{M}_1$ generates the data.
\citet{kmr+14} shows that the posterior distribution of $\alpha$
asymptotically concentrates around the value corresponding to the
true model and recommend the posterior median $\hat{\alpha}$ as the point estimate
for $\alpha$.
We adopt Beta$(0.5, 0.5)$ as the prior for the mixture weight $\alpha$, per
the recommendation of \citet{kmr+14}. Beta$(0.5, 0.5)$ equally encourages
the posterior density of $\alpha$ to concentrate around $0$ and $1$.
We also test the
sensitivity of our results to the prior on $\alpha$ by using
a uniform prior on $\alpha$.

\paragraph{Implementation}
We compute $\Delta_{\text{WAIC}}$ and its standard deviation
from the HMC posterior samples for $\mathcal{M}_1$ and $\mathcal{M}_2$.
Given the low dimensionality of the model, we are able to compute
the Bayes factor with the Sequential
Monte Carlo algorithm \citep{smc2, smc1} implemented in \texttt{pymc3}.
We implement the mixture model as a separate model in \texttt{pymc3} and 
sample from the posterior with the HMC algorithm to infer the mixture weight
$\alpha$. We obtain the median of the posterior distribution
of $\alpha$ and visually examine the posterior for concentration
of probability density around $0$ or $1$.
We present and interpret these model selection metrics in \S~\ref{sec:res-comp}.

\section{Results} \label{sec:res}

\begin{figure*}
    \gridline{\fig{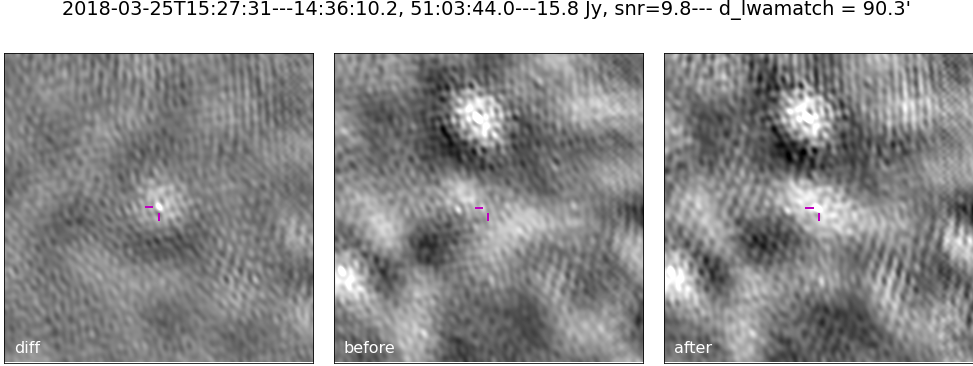}{0.6\textwidth}{(a)}}
    \gridline{\fig{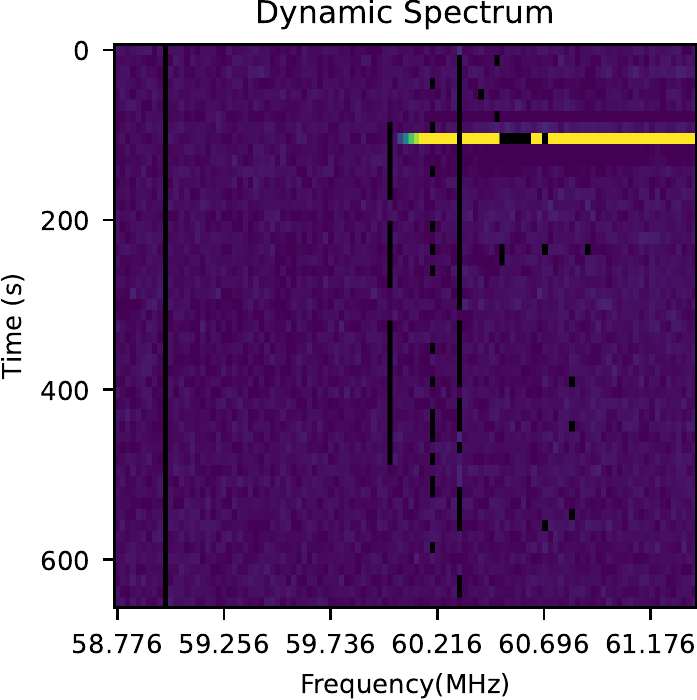}{0.3\textwidth}{(b)} \fig{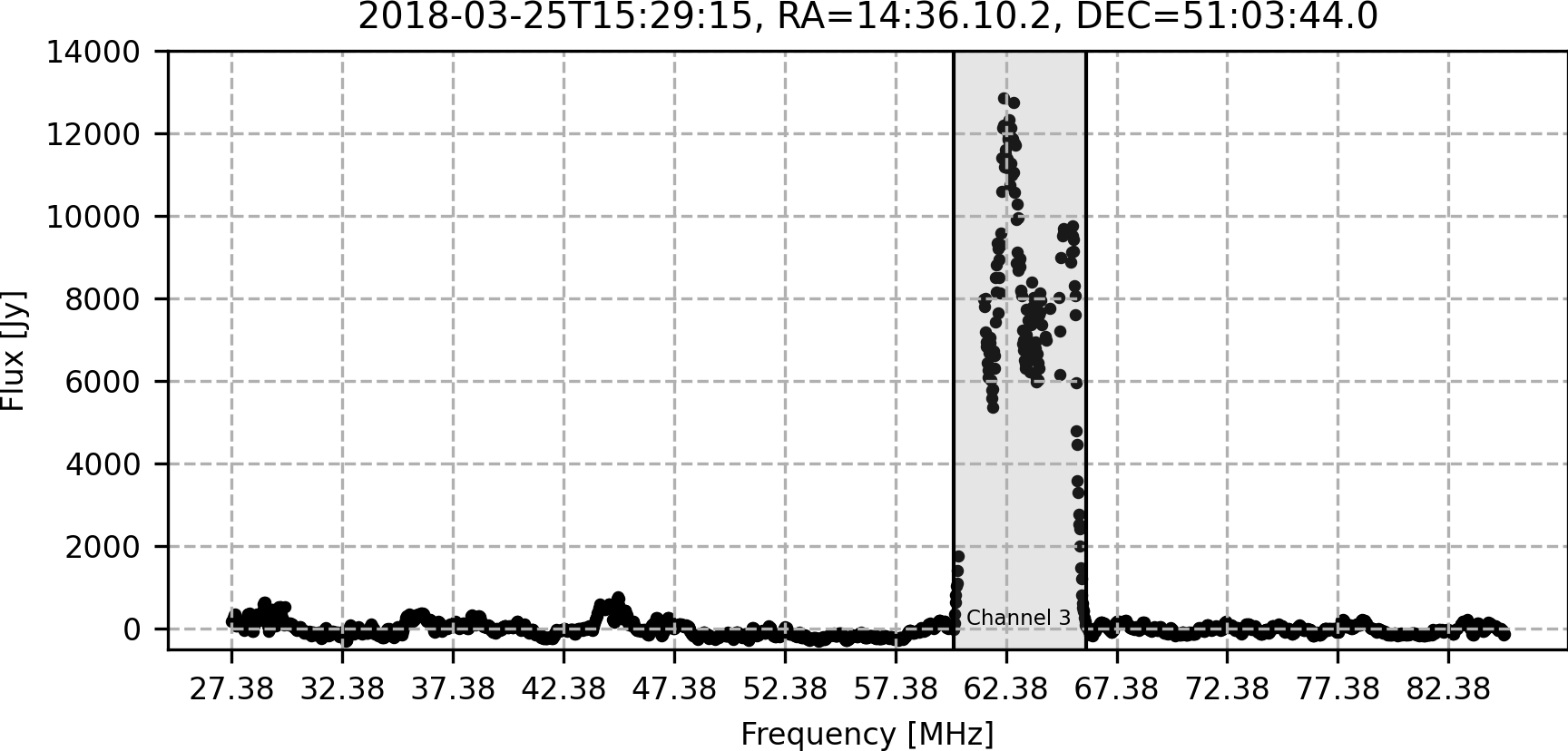}{0.6\textwidth}{(c)}}
    \caption{\label{fig:cand} Diagnostics of the unresolved reflection candidate OLWA~J1436+5103.
        \textbf{(a)} Discovery images of the candidate from the $722$~kHz wide search.
        The three panels show the differenced image, the image from the day before, and the image when the source
        appears. The title text displays the date of occurrence, the coordinates, the flux density, S/N, and distance
        to closest match in the persistent source catalog.
        \textbf{(b)} Dynamic spectrum for the $10$~min integration within a single $2.6$~MHz subband. The source
        is confined within a single time integration and only part of the subband bandwidth.
        \textbf{(c)} Spectrum of the source across the full $58$~MHz bandwidth in the single integration when
    the source is bright. The shaded region indicates broadcast frequencies of Channel 3 television.
    The coincidence of the emission frequencies with Channel 3 TV broadcast frequencies point to
    this source as a reflection artifact, likely from a meteor.}

\end{figure*}
\begin{figure}
    \epsscale{0.7}
    \plotone{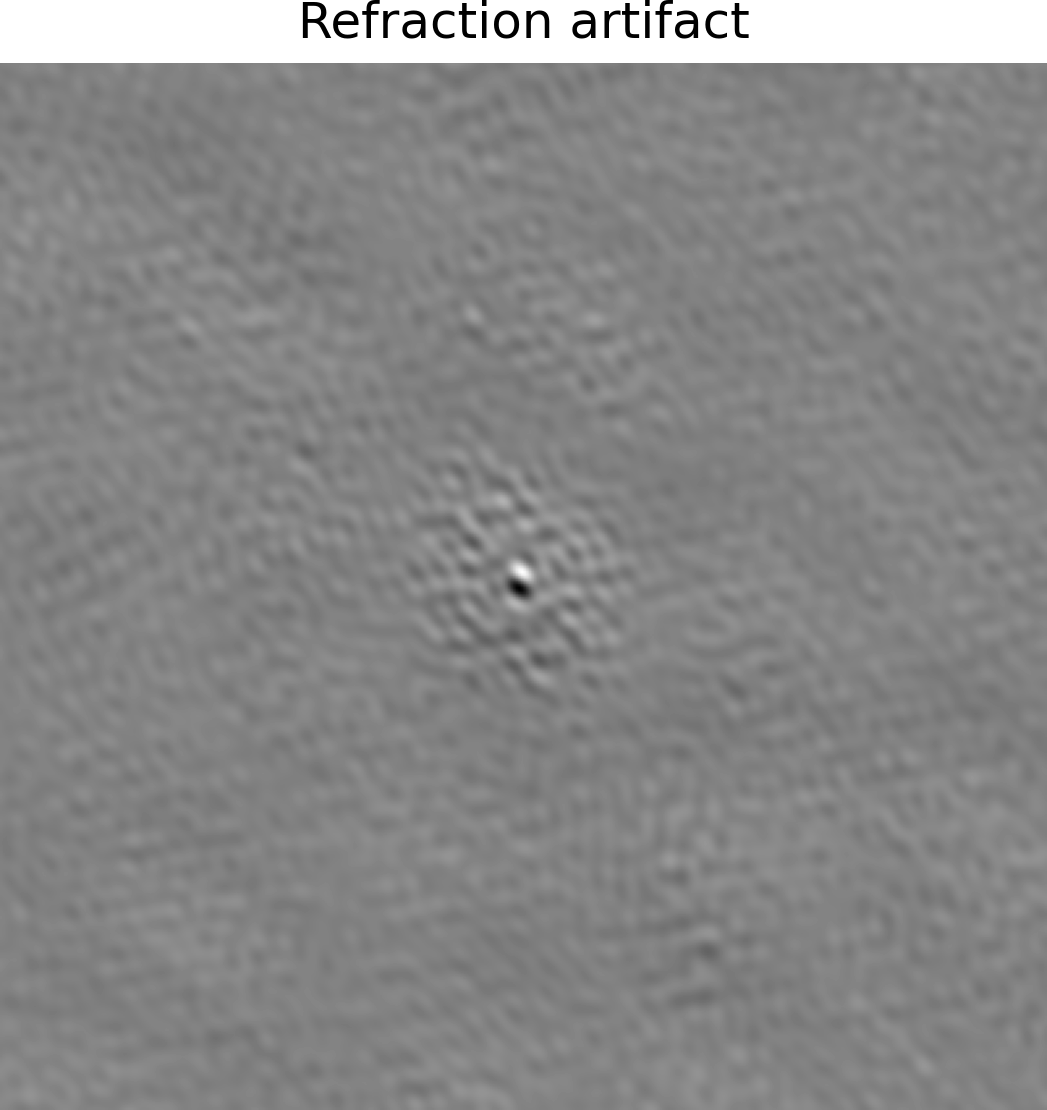}
    \caption{\label{fig:refraction}An example of refraction artifact in a differenced image. The position
    offset of the source between the two images gives rise to the dipole pattern in the differenced image.}
\end{figure}

\subsection{Artifacts}
\begin{deluxetable}{c c c c c}
\tablecolumns{2}
\tablecaption{Number of transient candidates remaining after each major vetting step of
the transient detection pipeline}
\label{tab:pipeline}
\tablehead{\colhead{Search step} &  \colhead{Detection count}}
\startdata
Source detection & 9057 \\
Persistent-source matching & 2317 \\
Visual inspection & 2\tablenotemark{a} \\
Re-imaging & 0
\enddata
\tablenotetext{a}{One of the two remaining candidate is a sidelobe of a scintillating Vir A and disappears after deconvolving Vir A.
The second candidate is the bright meteor reflection shown in Fig.~\ref{fig:cand}.}
\end{deluxetable}
Table~\ref{tab:pipeline} shows the number of transient candidates after each
sifting step. All $9057$ detected sources turned out to be artifacts.
All of the artifact classes detailed in \citet{ahe+19} appear in our data: 
meteor reflections, airplanes, horizon RFI sources, and scintillating sources. Fig.~\ref{fig:cand} shows
a bright meteor reflection candidate, which appears as an unresolved source in the
image. 
In addition to the artifacts detailed in \citet{ahe+19}, we identify $2$ classes of artifacts
that are unique to our sidereal differencing search with long integration time: refraction artifacts and spurious
point-like sources near the NCP.

The first class of artifacts that we identify is refraction artifacts
\citep[also described in][]{kle+07}.
The bulk ionosphere functions as a
spherical lens for a wide-field array \citep{vkb+14}. Due to the difference in the bulk
ionospheric content between two images that are $1$ day apart, sources are refracted by different amounts
in the two images and result in artifacts that have a dipole shape in the subtracted images
(see Fig.~\ref{fig:refraction} for an example).
We identify these artifacts by visual inspection and by cross-matching detections against the persistent
source catalog generated
as a by-product of \citet{ahe+19}. However, for more sensitive searches in the future, the number of refraction artifacts will increase;
collectively, their sidelobes may raise the noise level significantly.
Image-plane de-distortion techniques like \texttt{fits\_warp} \citep{hh18} and direct
measurement \& removal techniques \citep[see e.g.][]{r16} can be used to
suppress these refraction artifacts and their sidelobes in future searches, provided that
the ionospheric phase remains coherent across the array.

\begin{figure}
    \plotone{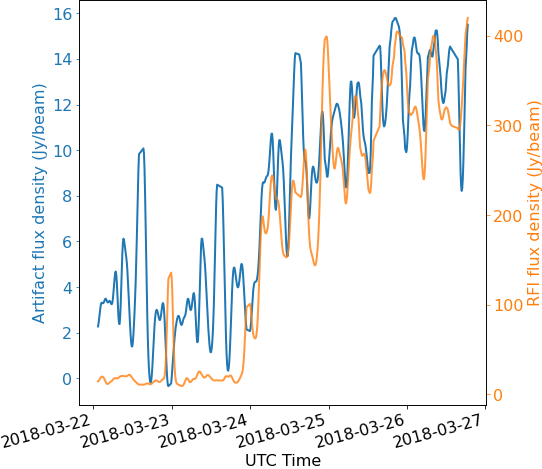}
    \caption{\label{fig:rfi-lc}Light curves of the point source artifact at $\delta=86^{\circ}$ and the horizon
    RFI source. The flux scale for the artifact is on the left vertical axis and the flux scale for the horizon RFI source on the
right. The light curves of these two sources are correlated.}
\end{figure}

The second class of artifacts is spurious point sources near the NCP.
Two prominent sources, one at $\delta=86^{\circ}$ and the other at $\delta=76^{\circ}$, were repeatedly
detected. Their flux density values correlate with that of a source of RFI in the northwest,
which we attribute to an arcing power line (Fig.~\ref{fig:rfi-lc}).
For a long integration time, the slow fringe rate near
the NCP may allow low-level near-field RFI sources and their sidelobes
to show up as point-like sources \citep{r02, obz+13}. For this reason, we exclude
the $15^{\circ}$ radius around the NCP from our subsequent analyses.

We note that even though the \citet{sfb+16} survey centered on the NCP and they did not test for
an RFI source outside their $10\deg$ FOV, it is unlikely that their detection is a sidelobe of a source of RFI.
Unlike the OVRO-LWA, which cross-correlates all dipole antennas, LOFAR first beamforms on the
station level (each station consisting of $96$ signal paths, typically 48 dual-polarization antennas)
and then cross-correlates voltages from different stations.
The station-based beamforming approach suppresses sensitivity to sources outside the main beam.
In addition, although all the individual LOFAR dipole antennas are aligned,
the antenna configurations of the Dutch LOFAR stations are rotated with respect to each other \citep{lofar}, making it even less likely
for the pair of stations in each baseline to be sensitive to the same direction far beyond the main beam.
Finally, deep LOFAR observations of the NCP did not reveal RFI artifacts \citep{obz+13b}.
Therefore, despite the high declination of the \citet{sfb+16} survey field, we conclude that the sidelobe of a horizon
RFI source likely did not lead to their transient detection.

\subsection{Limits on Transient Surface Density}

\begin{deluxetable}{c c}[htb]
\tablecolumns{2}
\tablecaption{Sky area per detection threshold bin at $10$~min timescale\label{tab:rho-s}}
\tablehead{
    \colhead{Detection threshold (Jy)} & \colhead{Sky area (deg$^2$)}
}
\startdata
     5.33 & 242.36 \\
     5.44 & 381.59 \\
     5.54 & 479.56 \\
     5.65 & 835.37 \\
     ... & ... \\
     58.07 & 14.1 \\
\enddata
\tablecomments{Table~\ref{tab:rho-s} is published in its entirety in the machine-readable format.
      A portion is shown here for guidance regarding its form and content.}
\end{deluxetable}
Fig.~\ref{fig:noises} illustrates the noise characteristics of the survey.
Across the survey, the mean noise level in subtracted images is $1.57$~Jy with a standard
deviation of $0.39$~Jy. Given our $6.5\sigma$ detection threshold, the mean noise level translates to a sensitivity of
$10$~Jy at zenith. The cumulative sky area surveyed as a function of sensitivity is shown in Fig.~\ref{fig:rho-s},
with the differential area per sensitivity bin recorded in Table.~\ref{tab:rho-s}.
As we find no astrophysical transient candidates in our search, we seek to put an upper limit in
the transient surface density-flux density phase space.
Our search is done with sidereal image differencing with an integration time of $10$~minutes.
The number of sidereally differenced $10$~min images $N$ (Eq.~\ref{eq:omegatot})
is $N=659$ after flagging integrations with excessive noise.

Because we exclude the sky area with declination above $75\deg$ and altitude angle below $30\deg$,
we calculate the snapshot FOV and the FOV-averaged sensitivity numerically.
We begin with a grid defined by 
the cosine of the zenith angle, $\cos\theta$,
and the azimuth angle, $\phi$, such that each grid cell has the same solid angle $\Omega$. 
We then exclude cells that do not satisfy our declination cut. Finally, we evaluate the total
solid angle integral $\Omega=\int\int (d\cos\theta) d\phi$ and the beam-averaging integral (Eq.~\ref{eq:sens})
by a Riemann sum over the remaining grid cells. We find that the effective snapshot FOV for our
survey is $\Omega_{FOV}=9800\deg^2$ and the FOV-averaged sensitivity is $1.7\sigma_z$.

Therefore, for a given population of transients with timescale $T$ from $10$ min to $1$ day,
the total sky area searched for a transient with timescale $T$ is
\begin{eqnarray}
    \Omega_{tot}&=& \left.\Omega_{FOV} N \middle/\left\lfloor\frac{T}{10\text{min}}\right\rfloor\right. \nonumber\\
    &=&\left. 6.5\times 10^6 \middle/\left\lfloor\frac{T}{10\text{min}}\right\rfloor \right. \deg^2.
\end{eqnarray}
We found no $10$~min transients at an averaged sensitivity of $S_0=17$~Jy. At this flux level,
we apply the approach described in \S~\ref{sec:freq} and place a $95\%$ confidence frequentist limit
on the transient surface density at
\begin{equation}
    \rho \leq 4.6 \times 10^{-7} \left\lceil\frac{T}{10\text{min}}\right\rceil \deg^{-2}.
\end{equation}
We place our limits in the context of other surveys at similar frequencies in
Fig.~\ref{fig:phasespace}.
Even though our upper limit is a factor of 30 more stringent than that of 
\citet{sfb+16}, our upper limit is marginally consistent with their 95\% confidence lower limit of
$1.5\times10^{-7}\deg^{-2}$ at $11$~min timescale and $15$~Jy.

We apply our Bayesian model $\mathcal{M}_1$ to the detection threshold-sky area data (Table.~\ref{tab:rho-s}).
The model jointly infers the flux density distribution power law index $\gamma$ and the reference
surface density
at $15$~Jy, $\rho_*$, because our survey probes different amount of
volume depending on $\gamma$. The estimate on $\rho_*$ is averaged over
the prior on $\gamma$.
In the uninformative prior case, the posterior distribution of $\rho_*$ is dominated
by the prior for much of the probability density because the data do not contain much
information. We report a $99.7\%$ credible
upper limit of $2.1\times 10^{-7}\deg^{-2}$,
at which point the posterior distribution has deviated from the prior
significantly.
In the case of a uniform prior over $(0,5)$ on $\gamma$
and flat prior on $\rho_*$,
we find a $95\%$ credible upper limit of $3.9\times10^{-7}\deg^{-2}$ and a $99.7\%$ credible
upper limit of $8.2\times 10^{-7}\deg^{-2}$.

\subsection{Consistency with Stewart et al. (2016)}
\label{sec:res-comp}
\begin{deluxetable*}{c c c c c}
\tablecolumns{5}
\tablecaption{Survey parameters of this work with comparisons to the previous OVRO-LWA survey \citep{ahe+19}
and \citet{sfb+16} at relevant timescale\label{tab:parms}}
\tablehead{
    & \colhead{This work} & \colhead{\cite{ahe+19}} & \colhead{\citet{sfb+16}}
}
\startdata
Timescale & 611 s -- 1 day  & 13 s -- 1 day & 30s, 2 min, 11 min\tablenotemark{a}, 55 min, 297 min \\
Central frequency (MHz) & $60$  & $56$ & $60$ \\
Bandwidth (kHz) & $744$  & $58000$ & $195$ \\
Resolution (arcmin) & $23\times 13$ & $29\times 13.5$ & $5.4\times 2.3$ \\
Total observing time (hours) & 137 & 31 & 348 \\
Snapshot FOV ($\deg^2$)& $9800$ & $17,045$ & $175$ \\
Average rms (Jy/beam) \tablenotemark{b} & $1.57$ & $1.68$ & $0.79$\tablenotemark{c} \\
95\% surface density upper limit ($\deg^{-2}$) \tablenotemark{d}& $4.6\times10^{-7}$ & $5.53\times10^{-7}$ & $1.4\times10^{-5}$ \\
95\% surface density lower limit ($\deg^{-2}$) \tablenotemark{d}& - & - & $1.5\times10^{-7}$\\
\enddata
\tablenotetext{a}{The search at this timescale yielded a detection.}
\tablenotetext{b}{Average rms is quoted at the $6$~min timescale for \citet{ahe+19} and the
    $11$~min timescale for \citet{sfb+16}, the timescales of interest in this work.}
\tablenotetext{c}{The detected transient had a flux density of $15$~Jy in a single integration,
but the flux density was suppressed in the detection image due to deconvolution artifacts.}
\tablenotetext{d}{Frequentist estimate.}
\end{deluxetable*}

Table~\ref{tab:parms} compares the parameters of our survey to \citet{sfb+16} and \citet{ahe+19}. 
Our survey features a similar bandwidth, sensitivity, and timescale as the transient ILT~J225347+862146.
We ask whether our results are consistent with the \citet{sfb+16} detection
in a Bayesian model comparison setting. We consider the \citet{sfb+16} detection as a data point
$D_{L}$ (Eq.~\ref{eq:dl}), and our survey as a collection of data points $\{D_{O,i}\}$ given by Table~\ref{tab:rho-s}.
Model $\mathcal{M}_1$ posits that both observations can be explained by a single population,
whereas $\mathcal{M}_2$ posits that our survey's selection effect results in a reduced transient rate
(or equivalently, that our survey probes a different population with a reduced surface density).
We consider the WAIC,
the Bayes factor $B_{12}$, and the mixture model
parameter $\alpha$ as three separate tests. We vary the prior on the surface density ratio $r$ and show the metrics in
Table.~\ref{tab:comp}.

\begin{deluxetable*}{L C C | C | C }
\tablecolumns{5}
\tablecaption{\label{tab:comp}Model comparison metrics between the single rate model, $\mathcal{M}_1$,
    and the two-rate model, $\mathcal{M}_2$, with different priors on the rate ratio $r$ for the OVRO-LWA
    survey. $\Delta{\text{WAIC}}_{12}$ is the difference in WAIC, $\sigma_{\Delta{\text{WAIC}},12}$ its uncertainty,
    $B_{12}$ the Bayes factor, and $\hat{\alpha}$ the posterior median of the mixture weight.
    In all cases we additionally bound $0<r<1$ due to our non-detection. Larger values of $\Delta{\text{WAIC}}_{12}$,
    $B_{12}$, and $\hat{\alpha}$ mean greater preference for $\mathcal{M}_2$ relative to $\mathcal{M}_1$.}
\tablehead{
 \colhead{Prior} & \multicolumn{2}{c|}{\text{Predictive Accuracy}} & \multicolumn{1}{c|}{\text{Bayes Factor}} & \multicolumn{1}{c}{\text{Mixture Model}}
}
\startdata
             & \Delta\text{WAIC}_{12}& \sigma_{\Delta\text{WAIC},12} & B_{12} & \hat{\alpha} \\
\hline
r\sim \text{Uniform}(0,1) & 1.6 & 1.3 & 3.53 & 0.78 \\
p(r)\propto 1/r & 4.0 & 3.1 & 28.8 & 0.97 \\
1/r\sim \text{Uniform}(1,2\times10^4) & 4.1 & 3.1 & 31.8  & 0.97
\enddata
\end{deluxetable*}

For all the priors we chose for $r$, the difference in WAIC, which estimates the predictive power of
each model, is comparable to its standard deviation estimated across all data.
The high standard error estimate is consistent with the fact that all but one data point,
the detection, contain very little information.
The WAIC test is therefore inconclusive.

We are able to compute the Bayes factor with good precision, as estimated from the results from multiple
parallel MCMC chains.
The Bayes factor gives the ratio of the posterior probability of each model. In our case
where we assume the prior probability on each model to be equal, the Bayes factor corresponds to the ratio of the
likelihoods of observing the data under each of the two models.
The only addition in model $\mathcal{M}_2$ compared to $\mathcal{M}_1$ is the surface density ratio
$r$ for our survey relative to \citet{sfb+16}.
We compute the Bayes factor for different prior distributions over $r$.
We rely on the scale suggested by \citet{bf1}, which categorizes the Bayes factor significance as
``not worth more than a bare mention'' ($0<\log(B_{12})<1/2$),
``substantial'' ($1/2<\log(B_{12})<1$),
``strong'' ($1<\log(B_{12})<2$),
and ``decisive'' ($\log(B_{12})>2$),
to interpret the Bayes factor $B_{12}$.
The uniform prior on $r$ model presents ``substantial'' evidence,
the uninformative prior model ``strong'', and uniform prior on $1/r$ model ``strong'' evidence
that $\mathcal{M}_2$ is preferred.
Although the Bayes factor varies by up to an order of magnitude with the choice of prior,
in all cases the Bayes factor prefers $\mathcal{M}_2$.
Therefore, we conclude that the Bayes factor test prefers
the two-population model, $\mathcal{M}_2$.

\begin{figure}
    \epsscale{0.8}
    \plotone{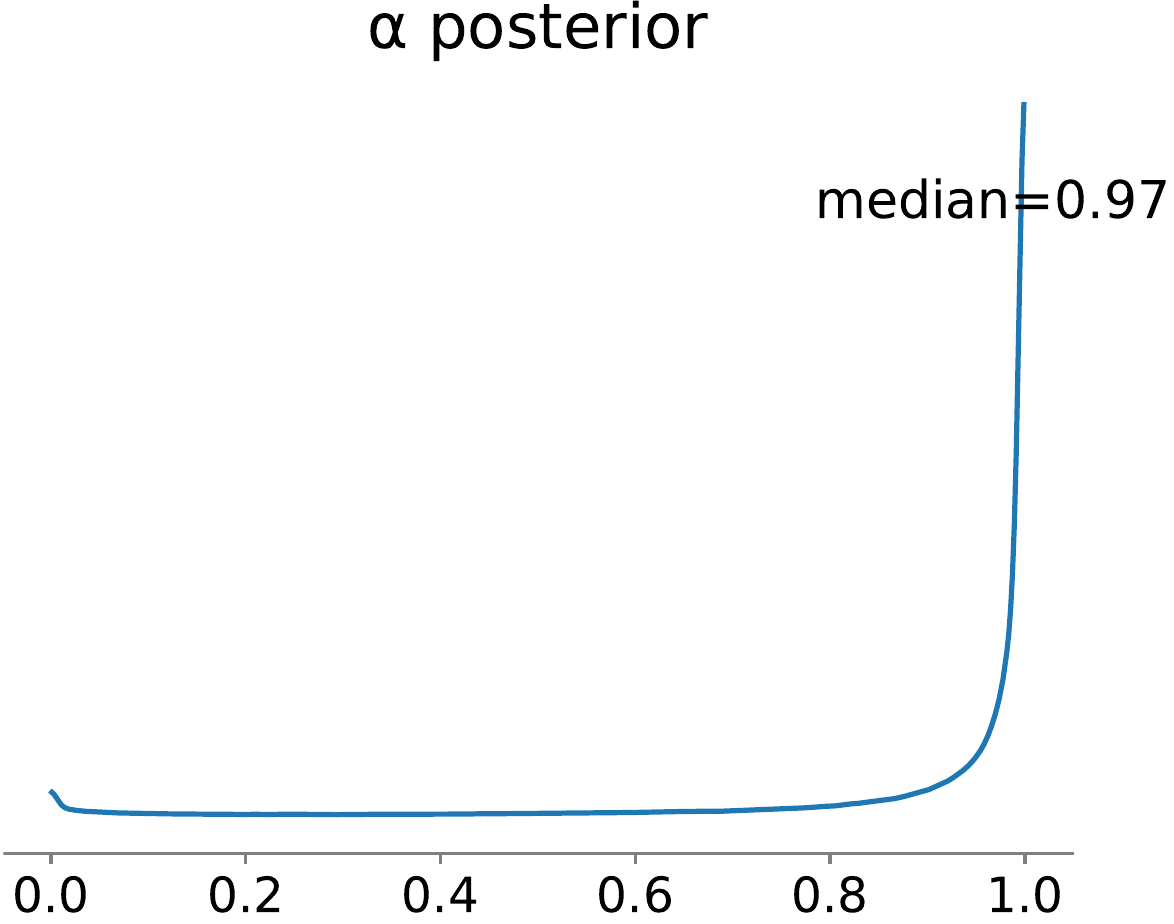}
    \caption{\label{fig:post} Posterior distribution of the mixture weight $\alpha$
        with uninformative prior on all parameters. We adopt 
        the posterior median $0.94$ to be the point estimate for $\alpha$.
        The posterior concentrates toward $\alpha=1$, indicating a
    preference for the model $\mathcal{M}_2$.}
\end{figure}

The mixture weight $\alpha$ tells a similar story as the Bayes factor.
Fig.~\ref{fig:post} shows a sample posterior distribution of $\alpha$.
For all of the $\mathcal{M}_2$ variants, the posterior distribution of $\alpha$ concentrates toward $1$,
exhibiting a preference for $\mathcal{M}_2$ \citep{kmr+14}. 
All of the posterior median estimates for $\alpha$, $\hat{\alpha}$ are close to $1$.
We draw identical conclusions in the case when the prior on $\alpha$ is uniform as well, but
only show results for the prior $\alpha\sim$Beta$(0.5,0.5)$.

In the tests that are conclusive, we find strong evidence in support of the model $\mathcal{M}_2$,
suggesting that our non-detection is not consistent with
\citet{sfb+16} under a single Poisson population model.
Since we did not have a detection, our goal for testing survey result
consistency is to inform designs for future surveys aiming to uncover this
population. The degree to which the statistical evidence are in favor of the two-population
model, $\mathcal{M}_2$, prompts us to consider why our survey may be inconsistent 
with \citet{sfb+16}.
Because our survey is narrow band and at comparable sensitivity, the only remaining
non-trivial differences between our survey and that of \citet{sfb+16} are the choice of
survey field and the time sampling. We consider how these differences may
explain the inconsistency and their implications on future survey strategies in \S~\ref{sec:disc}.

\section{Discussion} \label{sec:disc}
Motivated by the hypothesis that the \citet{sfb+16} transient, ILT~J225347+862146, may be narrowband,
we searched for narrowband transients in $137$ hours of all-sky data with the OVRO-LWA at matching
timescale and sensitivity as ILT~J225347+862146. Having searched almost two orders of magnitude larger
sky area for a $10$~min timescale transient than did \citet{sfb+16}, we did not detect any transient.
Using a collection of Bayesian model comparison approaches, we found compelling evidence that our
non-detection is inconsistent with \citet{sfb+16}. We discuss the implications
of our non-detection followed by details of an M dwarf coincident with ILT~J225347+862146 in this section.

\subsection{Implications of Our Non-detection}
Despite matching the \citet{sfb+16} survey as much as possible while
searching a much larger sky area, we did not detect any transient.
We also find compelling statistical evidence that our survey results are inconsistent with
that of \citet{sfb+16} under a single Poisson transient population model.
Assuming that the transient is astrophysical, we are left with two classes
of possibilities. First, \citet{sfb+16} may have been an instance of discovery bias.
Second, the remaining differences in survey design may have led to our non-detection.
We explore each of these scenarios and their implications on future surveys
aiming at unveiling the population associated with ILT~J225347+862146.

\subsubsection{Was It Discovery Bias?}
\label{sec:luck}
Perhaps the conceptually simplest solution for reconciling the \cite{sfb+16}
results with subsequent non-detections is that they found a rare instance of
the population \citep[see e.g.][for a discussion of the discovery bias at the population level]{me18}.
One such recent example is the first discovered Fast Radio Burst, the ``Lorimer burst'' \citep{lbm+07}.
The inferred
rate from the Lorimer burst for events with similar fluence ($\sim150$~Jy~ms) was
$400$~sky$^{-1}$~day$^{-1}$. However, subsequent searches at similar frequencies
but much greater FOV yielded an estimate of $\sim10\pm4$~sky$^{-1}$~day$^{-1}$
for events with fluence greater than $100$~Jy~ms \citep{smb+18}. 
To estimate how lucky \citet{sfb+16} was if our survey and theirs truly probe the
same population, we integrate the probability of obtaining a detection with a survey
like \citet{sfb+16}, $(1-P_{pois}(n=0|\lambda=\rho_*\Omega_{tot,L}))$,
over the marginal posterior distribution of the surface density at $15$~Jy,
$\rho_*$, inferred from our data $D_{O}$. This
probability turns out to be $0.0018$ under the uninformative prior and $0.02$
under the uniform prior.

On a technical note, previous surveys have quantified luck by calculating the null-detection
probability assuming a fixed $\gamma$ and using either the frequentist point estimate \citep[e.g.][]{kws+21}
or the $95\%$ confidence interval \citep[e.g.][]{ahe+19} from the detection.
The use of point estimate does not account for the significant uncertainty in the parameter,
whereas the use of the confidence interval does not capitalize on the fact that the
detection probability decays very quickly as $\lambda$ approaches $0$.
Because it integrates over the posteriors of both $\gamma$ and $\rho_*$, our estimate
of luck uses all the information available and makes minimal assumptions.

The detection probability that we calculated suggests that it is still
plausible that the \citet{sfb+16}
has been a very lucky incident and the event is a extreme outlier of
the fluence distribution.
Curiously, although the \citet{sfb+16} survey ran for about $4$ months, the transient
was detected on the first day of the survey, within the first $30$ $11$~min snapshots
taken.
Using the single population
model $\mathcal{M}_1$ with an uninformative prior,
combining our non-detection with the \citet{sfb+16} detection
yields a $95\%$ credible interval for the surface density $\rho_*$ of
$(3.5\times10^{-12}, 3.4\times10^{-7})\deg^{-2}$ and a point estimate of
$1.1\times10^{-7}\deg^{-2}$. In comparison, the surface density point estimate
implied by the \citet{sfb+16} detection is $2.9\times 10^{-6}\deg^{-2}$.
If we are indeed probing the same population
as \citet{sfb+16}, our non-detection establishes that the population associated
with their detection is much rarer than their detection has implied.

Future surveys that
aim at finding this transient will likely have diminishing returns, because
the population can be many orders of magnitude rarer than the \citet{sfb+16}
detection implied.
    The best effort to uncover the population associated with ILT~J225347+862146
in this case coincides with the systematic exploration of the low-frequency
transient phase space.
Future surveys will have to reach orders of magnitude better sensitivity,
run for orders of magnitude longer time period, and ideally use more
optimized time-frequency filtering in order to make significant
progress uncovering transients in the low-frequency radio transient sky. 
The Stage III expansion of the OVRO-LWA, scheduled to start observing in
early 2022, will feature redesigned analog
electronics that suppress the coupling in adjacent signal paths that limit our current
sensitivity. With the Stage III array, the thermal noise in a subtracted image
across the full bandwidth on $10$~min timescale will be $30$~mJy.
The processing infrastructure developed in this work and elsewhere \citep[see e.g.][]{rkr+21}
represent significant steps toward turning low-frequency radio interferometers
into real-time transient factories.

\subsubsection{Was It Selection Effects?}
On the other hand, the model comparison results compel us to consider the more likely
 scenario that
that our survey design has not selected for the same population as did \citet{sfb+16}.
While there is only one detection, our Bayesian approach did account
for the uncertainty that comes with the dearth of informative by drawing conclusion from
the full posterior distribution.
Our survey searched for narrowband transients, as did \citet{sfb+16}.
The only remaining substantial differences between our survey and \citet{sfb+16} are their choice of the NCP
as the monitoring field and their time sampling, spreading $400$ hours of observing time
over the course of $4$ months. We seek hypotheses that involve these two differences and not
luck.

First, we consider the possibility that the choice of NCP as the monitoring field made \citet{sfb+16} much
more likely than us to detect an instance of the population.
For an extragalactic population of transients, the events distribution should be isotropic.
If the transient population is galactic, the events should concentrate along the galactic plane.
If the distance scale of the population is less than the galactic scale height of $<400$~pc, the events
will appear uniform over the sky. If the distance scale of the population is much greater
than the galactic scale height, the events will concentrate at low galactic latitudes.
ILT~J225347+862146 has a galactic latitude of $b=28.6\deg$.
Finally, if a population of transients uniformly distributes across the sky,
but there is a bias against
finding sources at low Galactic latitudes, then the observed population may 
concentrate around high Galactic latitudes. Most of the sky area that our
survey probes is in high Galactic latitudes.
Thus, no populations of astrophysical transients
should concentrate only around the NCP when a sufficient depth is probed.
The NCP preference can only
be due to a extremely nearby progenitor relative to the rest of the population.
The NCP hypothesis requires \citet{sfb+16} again to be lucky, the consequences of which we already
discussed in \S~\ref{sec:luck}.

The other possibility, which ascribes less luck to \citet{sfb+16}, is that the difference in time
sampling between our survey and that of \citet{sfb+16} led to our non-detection.
Our survey consisted of $137$ hours of continuous observations, whereas \citet{sfb+16} monitored the NCP
intermittently over the course of $4$ months, totaling $\sim400$ hours of observations. 
Under a Poisson model, the cadence of observations, as long as it is much greater than the timescale
of the transient, does not affect the distribution of the outcome. So a population that is sensitive
to sampling cadence will necessarily have a non-Poisson temporal behavior.
We explore one simple scenario here with an order-of-magnitude estimation.
Over the timescale of years, suppose there is a constant number of sources in the sky
capable of producing this class of transients detectable by \citet{sfb+16}. Assuming that
\citet{sfb+16} was unaffected by the time clustering behavior of the bursts,
we take the mean surface density $\rho=0.006\deg^{-2}$, and the mean burst rate $r=0.003$~hr$^{-1}$,
from the FOV and total observing time of \citet{sfb+16}.
We take their point estimate of surface density and extrapolate that there are
$60$ such sources accessible to our survey based on our snapshot FOV.
In order for the probability of our observation falling outside any source's activity window
to be $>68\%$, the probability of
non-detection for an average individual source should be $>0.68^{1/60}=0.994.$
If we consider a model, where each source turns on for a short window $w$, emitting
bursts at roughly the observed burst rate by \citet{sfb+16}, then turns off for
a much longer time that averages around $T$, $T\gg w$.
Our non-detections can be readily realized if
the repeating timescale of the source $T>137\text{hr}/0.006\sim10^3$~days.
Stellar activity cycles or binary orbital periods can potentially give rise to these timescales. 
In contrast,
the $4$~month time-span of \citet{sfb+16} has probability $120/10^3=0.1$ of hitting the activity window.
This estimate still requires \citet{sfb+16} to be somewhat lucky and number of sources
in the sky to be few, but we do note
that there is significant uncertainty associated with this estimate.
Assuming that ILT~J225347+862146 is a
typical member of this population that produce temporally clustered bursts,
because the OVRO-LWA has a factor
of $50$ larger field of view, we can readily test this hypothesis by spreading
$\sim100$ hours of observations over the course of $\sim20$ days.
Although the added complexity of this explanation only made our non-detection
slightly more consistent with \citet{sfb+16}, the test for it is straightforward.

In summary, we have two remaining viable hypotheses. First, the \citet{sfb+16} detection
may represent an extreme sample of the fluence distribution, in which case more
sensitive and longer surveys may uncover the population. However, improving survey
sensitivity and duration has diminishing return if one's sole goal is to detect members of
this population,
since the surface density and the
fluence distribution power law index of the population cannot be well constrained from
existing observations \citep[see also][]{kipping21}.
It is however likely that the population will eventually be
revealed as low-frequency transient surveys becomes more sensitive and more automated.
The other hypothesis,
that the population are clustered in time, can be readily tested by spacing out observing
time with a wide-field instrument like the OVRO-LWA and AARTFAAC \citep{aartfaac}.

A potential alternative to our phenomenological approach for inferring the properties
of this class of transients is population synthesis
\citep[see e.g.][]{blrs14, glcp19} for potential progenitors. However, the
significant uncertainty associated with the single detection will likely give
inconclusive results.

\subsubsection{Limitations}
Two limitations may hinder our ability to understand the population underlying
ILT~J225347+862146 with our survey: unoptimized matched filtering for the population,
and incomplete characterization of survey sensitivity.

Although our choice of integration time and bandwidth is well-matched to
the event ILT~J225347+862146, our choice may not be well-matched to the
population of transients underlying ILT~J225347+862146.
It is possible that the
population has widely-varying timescales and frequency structures that our 
survey is not optimized for.
Even if our filtering is well matched to the typical timescales and frequencies,
because our $10$~min integrations do not overlap, we may
miss transients that do not fall entirely in a time integration.
However, because our FOV is much greater than that of \citet{sfb+16}
and these features are common to both our survey and that of \citet{sfb+16},
filtering mismatch for the population alone cannot explain our non-detection and does
not alter the implications of our results.
We only searched around 60~MHz in order to replicate the \citet{sfb+16} survey as much as possible,
but the transient population should manifest at other similar frequencies as well.
To maximize the chance of detecting a transient,
    a future transient survey with the OVRO-LWA may 
    feature overlapping integrations, overlapping search frequency windows, and different
search bandwidths across the $>57$~MHz observing bandwidth.

We quantified our sensitivity in terms of the rms of the subtracted image and assume
that our search is complete down to the detection threshold. Although we do routinely detect
refraction artifacts down to our detection threshold and we exclude regions in the sky that
are artifact-prone, the most robust way to assess completeness is via injection-recovery tests
that cover different observing time, elevation angles, and positions in the sky.
The completeness function over flux density can then be incorporated into our Bayesian rate
inference model.

\subsection{An M Dwarf Coincident with ILT~J225347+862146}
\begin{figure}
    \epsscale{1.2}
    \plotone{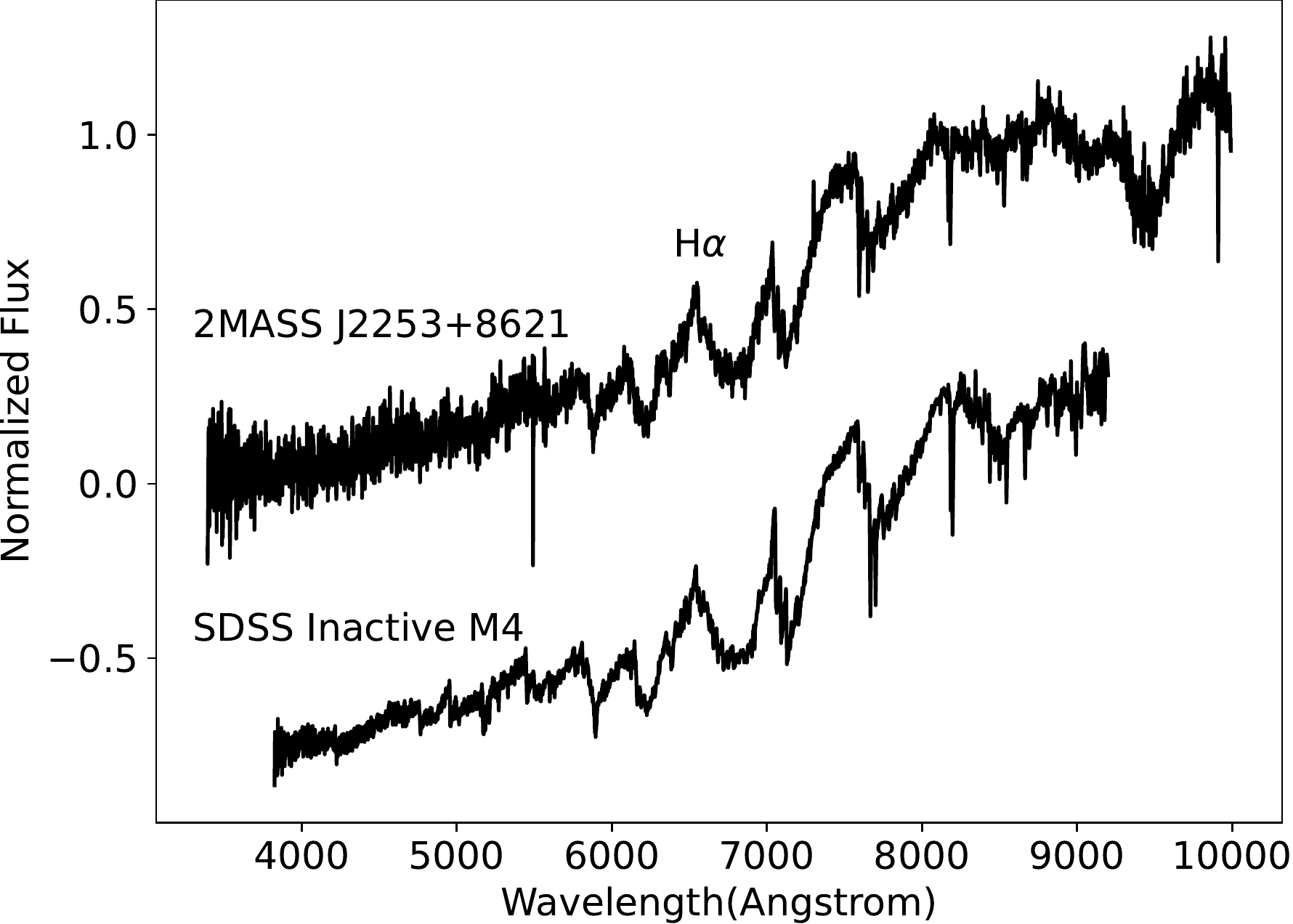}
    \caption{\label{fig:dbspspec}Palomar DBSP spectrum of the M dwarf 2MASS~J22535150+8621556
        coincident with the radio transient ILT~J225347+862146. The location of the 6562 ~\r{A}
        H$\alpha$ line is indicated.
An SDSS inactive M4 dwarf template spectrum \citep{bwh+07} is plotted with offset for reference.
The feature at $7300$~\r{A} was present in other sources during the same night of observation and is
thus likely not astrophysical.}
\end{figure}
\begin{deluxetable}{c c }
\tablecolumns{2}
\tablecaption{Basic parameters for the coincident M dwarf}
\label{tab:mdwarf}
\tablehead{Parameter & Value}
\startdata
2MASS Designation & 2MASS~J22535150+8621556\tablenotemark{a} \\
{\it Gaia} Designation & Gaia~EDR3~2301292714713394688\tablenotemark{b} \\
Right Ascension (J2000) & $22^h53^m51.45^s$ \\
Declination (J2000) & $+86^{\circ}21'55.56''$\\
Distance & $420^{+18}_{-22}$~pc\tablenotemark{c}\\ 
{\it Gaia} G magnitude & 18.8\tablenotemark{b} \\
{\it Gaia} Bp-Rp color & 2.59\tablenotemark{b} \\
Spectral type & M4V\\
\enddata
\tablenotemark{a}{2MASS \citep{2mass}}
\tablenotemark{b}{{\it Gaia} EDR 3 \citep{edr3}}
\tablenotetext{c}{{\it Gaia} EDR 3 geometric distance \citep{brf+21}}
\end{deluxetable}
Without a detection of another instance of the transient population,
we revisit an optical coincidence of the \citet{sfb+16} transient
for clues on the nature of the population.
In an attempt to elucidate the nature of ILT~J225347+862146,
\citet{sfb+16} obtained a deep ($r'\sim 22.5$) image of the field.
There was no discernible galaxy in their image.
For a galactic origin, \citet{sfb+16} considered radio flare stars,
in particular M dwarfs, as viable progenitors to
this population of transients.
In their optical image, they found one high-proper-motion objects
within the $1\sigma$ localization circle. They concluded that the object did not have
colors consistent with an M dwarf, noting however that their color calibration had significant errors.

We cross-matched the $1\sigma$-radius localization region of
ILT~J225347+862146 with the {\it Gaia} \citep{gaia} Early Data Release 3 source
catalog \citep{edr3} and found two matches.
The closer match, at an offset
of $10\arcsec$, is an M dwarf at a distance of $420^{+18}_{-22}$~pc \citep{brf+21}.
The M dwarf is indeed the high-proper-motion object identified by \citet{sfb+16}.
The farther offset match at $13\arcsec$ is a K dwarf at a distance of $1.7 \pm 0.2$~kpc \citep{brf+21}.

In order to prioritize follow-up efforts, we used the procedures outlined below to evaluate the
significance of the coincidence and attempted to identify a posteriori bias.
We did not seek to claim an association of the star with the transient in this exercise.
Rather, we assessed whether the coincidence warranted further investigations into any of these objects.
We emphasize that only more instances of the population, or observed peculiarities of the coincident
stars that may explain the transient, can lend credence to the association claim of the transient with a stellar source.

For each object, we randomly selected locations in the \citet{sfb+16} survey field
and searched for objects with parallax greater than the
$1\sigma$ upper bound of the object within the $1\sigma$ localization radius of $14\arcsec$ and
calculated the fraction of trials that resulted in matches. The calculated fraction represented the
chance of finding any object within the $14\arcsec$ localization radius with greater parallax
than the match in question.
We found this chance coincidence probability to be $1.9\%$ for
the M dwarf and $15\%$ for the K dwarf.
The probability of finding any galactic {\it Gaia} source within a $14\arcsec$ radius in
the \citet{sfb+16} field is $16\%$.
We used distance as a discriminating factor because bright transients from a nearer
source is in general energetically more plausible.
The low chance association rate is not due to survey incompleteness for dim sources, because
{\it Gaia} is $>99\%$ complete down to $G > 20$ at this declination \citep{be20}.
Although our chance coincidence criteria were quite general,
the criteria were determined \textit{after} the we identified the
coincidence. As such, the significance of the coincidence may be inflated. 
Based on the low chance coincidence rate, we decided to obtain
follow-up data on the M dwarf.

We obtained a spectrum of the M dwarf with the Double Spectrograph \citep[DBSP;][]{dbsp}
on the 200-inch Hale telescope.
The spectrum is consistent with an inactive M4 dwarf, exhibiting no excess H$\alpha$ emission nor
signs of a companion.
The {\it Gaia} \citep{edr3}, Wide-field Infrared Survey Explorer \citep[WISE;][]{wise},
and Two Micron All Sky Survey \citep[2MASS][]{2mass} colors are consistent
with a main sequence M4 dwarf.
Table~\ref{tab:mdwarf} summarizes the basic properties of the M dwarf.
We searched for signs of variability in other wavelengths.
The M dwarf was marginally detected in the Transiting Exoplanet Survey Satellite \citep[TESS;][]{tess}
Full Frame Images (FFIs) for sectors 18, 19, 20 as well as
Zwicky Transient Facility \citep[ZTF;][]{ztf} Data Release 6, and not detected in
Monitor of All-sky X-ray Image \citep[MAXI;][]{maxi}.
The light curves from TESS\footnote{generated with simple aperture photometry from the FFIs with
the package \texttt{lightkurve} \citep{lightkurve}}, ZTF, or MAXI did not show any transient behavior,
with the caveat of low signal-to-noise ratios.

If the M dwarf was responsible for the transient, the implied peak isotropic spectral luminosity
$L_{\nu}\sim3\times10^{21}$~erg~Hz$^{-1}$s$^{-1}$. The peak luminosity of the transient, assuming that
the emission is broadband, is $\nu L_{\nu}\sim2\times10^{29}$~erg~s$^{-1}$. 
The peak luminosity and the peak spectral luminosity would be many orders of magnitude higher than
those of the brightest bursts ever seen
from stars at centimeter to decameter wavelengths \citep[e.g.][although they were both targeted observations]{sm76, ob08}.
Given the lack of observed peculiarity of the M dwarf, we are unable to ascertain its association
with the transient.
\section{Conclusion} \label{sec:conc}
We presented results from a $137$ hr transient survey with the OVRO-LWA.
We designed the survey to search in a narrow bandwidth, in a much greater sky area,
and with enough sensitivity to detect
events like the low-frequency transient ILT~J225347+862146 discovered by \citet{sfb+16}.
We also presented an M dwarf coincident with this transient and optical
follow-up observations. This work represents the most targeted effort to date to
elucidate the nature of the population underlying this transient.
The main findings of this work are as follows:
\begin{enumerate}[noitemsep]
    \item We adopted
        a Bayesian inference and model comparison approach to model and compare transient surveys.
Our Bayesian approach accounts for our widely varying sensitivity as a function of FOV and
different transient population properties.
It can be extended readily to model the nuances of each transient survey.

    \item Despite searching
        for almost two orders of magnitude larger total sky area, our narrowband
transient search yielded no detections.
One possible explanation for our non-detection
and the non-detection of the \citet{ahe+19} broadband search
is that \citet{sfb+16} detected an extreme sample of the fluence distribution
(i.e. discovery bias). In
this scenario, we revised the surface density of transients like ILT~J225347+862146 to
$1.1\times 10^{-7}\deg^{-2}$, a factor of 30 
lower than the estimate implied by the \citet{sfb+16} detection.
The $95\%$ credible interval of the surface density is
$(3.5\times 10^{-12}, 3.4\times10^{-7})\deg^{-2}$, 
\item The alternative explanation is that the population produces transients that are clustered in time
with very low duty cycles and low all-sky source density.
Therefore, compared to the $4$ month time baseline of
\citet{sfb+16}, our short time baseline ($5$ days) was responsible for our non-detection.
Because our much larger FOV compared to \citet{sfb+16}, the allowed parameter space for this hypothesis
is small.
However, the cost for testing this hypothesis is relatively low.

\item Owing to the availability of the {\it Gaia} catalog, we identified an object within the
$1\sigma$ localization region of ILT~J225347+862146 as an M dwarf at $420$~pc, with
an a posteriori chance coincidence rate $<2\%$.
However, we are unable to robustly associate
this M dwarf with the transient based on follow-up spectroscopy and existing catalog data.
\end{enumerate}

\begin{acknowledgments}
We thank the anonymous referee for a thoughtful report that improved the quality
of the manuscript.
We thank Casey Law for a thorough reading of the manuscript.
We are indebted to Viraj Karambelkar, Mansi Kasliwal, Andy Tzanidakis,
Yuhan Yao, and the ZTF team for the DBSP observation and data reduction.
We thank Barak Zackay, Yuhan Yao, Sarah Blunt, and Ryan Rubenzahl for helpful discussions.

This material is based in part upon work
supported by the National Science Foundation under Grant Nos.
AST-1654815, AST-1212226, and AST-1828784.
This work was supported by a grant from the Simons Foundation (668346, JPG).
We are grateful to Schmidt Futures for supporting the Radio Camera Initiative,
under which part of this work was carried out.
The OVRO-LWA
project was initiated through the kind donation of Deborah
Castleman and Harold Rosen.
Y.H. thanks the LSSTC Data Science Fellowship
Program, which is funded by LSSTC, NSF Cybertraining Grant \#1829740, the Brinson Foundation,
and the Moore Foundation; his participation in the program has benefited this work.
G.H. acknowledges the
support of the Alfred P. Sloan Foundation and the Research
Corporation for Science Advancement.

This work has made use of data from the European Space Agency (ESA) mission
{\it Gaia} (\url{https://www.cosmos.esa.int/gaia}), processed by the {\it Gaia}
Data Processing and Analysis Consortium (DPAC,
\url{https://www.cosmos.esa.int/web/gaia/dpac/consortium}). Funding for the DPAC
has been provided by national institutions, in particular the institutions
participating in the {\it Gaia} Multilateral Agreement.
This research has made use of the NASA/IPAC Extragalactic Database, which is funded by the National
Aeronautics and Space Administration and operated by the California Institute of Technology.
This publication makes use of data products from the Two Micron All Sky Survey, which is a
joint project of the University of Massachusetts and the Infrared Processing and Analysis
Center/California Institute of Technology, funded by the National Aeronautics and
Space Administration and the National Science Foundation.
This publication makes use of data products from the Wide-field Infrared Survey Explorer, which
is a joint project of the University of California, Los Angeles, and the Jet Propulsion
Laboratory/California Institute of Technology, funded by the National Aeronautics and Space
Administration. This research has made use of the SIMBAD database,
operated at CDS, Strasbourg, France.
\end{acknowledgments}

\facilities{Hale~(DBSP), Gaia}

\software{astropy \citep{2018AJ....156..123A}, TTCal \citep{TTCal}, WSClean \citep{wsclean},
    CASA \citep{casa}, CASA~6 \citep{casa6},
    seaborn \citep{seaborn}, Jupyter \citep{jupyter}, Matplotlib \citep{matplotlib},
    pymc3 \citep{pymc3}, arviz \citep{arviz}, lightkurve \citep{lightkurve}, pyraf-dbsp \citep{pyraf-dbsp},
    and
    SciPy \citep{2020Scipy-NMeth}}

\bibliography{refs}{}
\bibliographystyle{aasjournal}

\appendix

\section{Derivation of an Uninformative Prior}
\label{sec:prior}
When surveys contain very few detections, the choice of prior can
impact the results of the inference quite significantly. Here we
derive a prior on our model parameters that is less informative
than a uniform prior.
We write our model in simplified notations as
\begin{equation}
    \lambda = \rho_* S^{-\gamma},
\end{equation}
where $\lambda/\Omega_{tot}\rightarrow \lambda$, $S/S_*\rightarrow S$
when compared to Eq.~\ref{eq:lamb}. We seek to derive a prior distribution
density function $p(\rho_*, \gamma)$ that is invariant under
reasonable reparameterization, such that it does not encode
information based on the parameterization of the problem.
Here we follow \citet{j46} and \citet{v14} and
derive one such prior using the symmetry of the model under exchange of
variables.
Since $S$ and $\lambda$ are symmetric in this relationship, the model
can also be rewritten as
\begin{equation}
    S = \rho_*' \lambda^{-\gamma'},
\end{equation}
i.e. a model of typical flux density changing with occurrence rate.
We can solve for the transformation
$\rho_*'=\rho_*^{1/\gamma}$ and $\gamma'=1/\gamma$.

The prior density function transforms as follows
\begin{equation}
    p(\rho_*, \gamma) d\rho_* d\gamma = q(\rho_*', \gamma')d\rho_*' d\gamma', 
\end{equation}
where $q(\rho_*', \gamma')$ is the prior density function on the reparameterized parameters.
Because we claim the same ignorance whether we parameterize the problem with
$(\rho_*, \gamma)$ or $(\rho_*',\gamma')$, the prior distribution function on the
two parameterization must be the same:
\begin{equation}
    p(\rho_*,\gamma_*) = q(\rho_*', \gamma').
\end{equation}

The determinant of the Jacobian matrix of the transformation
$(\rho_*, \gamma)\rightarrow(\rho_*', \gamma')$ is $-\rho^{\frac{1}{\gamma}-1}/\gamma^3$.

The change of variable theorem then gives
\begin{equation}
    p(\rho_*, \gamma) d\rho_* d\gamma = \left|-\frac{\rho^{\frac{1}{\gamma}-1}}{\gamma^3}\right| p(\rho_*^{1/\gamma}, 1/\gamma)d\rho_* d\gamma. 
\end{equation}
Imposing that the $\rho_*$ and $\gamma$ are independent in our prior, a functional form that
satisfies the above requirement is
\begin{eqnarray}
    p(\rho_*) \propto 1/\rho_*, \label{eq:prho}\\
    p(\gamma)\propto\gamma^{-3/2}.
\end{eqnarray}
When we modify $\rho$ to $r\rho$ 
in the two-population model $\mathcal{M}_2$ (Eq.~\ref{eq:m2}), Eq.~\ref{eq:prho} is satisfied when $p(r)\propto 1/r$. This prior density is also
invariant under the reparameterization $r\rightarrow 1/r$.
\end{CJK*}
\end{document}